\documentclass{aa}  

\usepackage[utf8]{inputenc}
\usepackage[T1]{fontenc}

\usepackage{natbib}
\bibpunct{(}{)}{;}{a}{}{,}
\newcommand{\citeay}[1]{\citeauthor{#1} \citeyear{#1}}

\usepackage{graphicx}
\usepackage[font=small, labelfont=bf, justification=justified, singlelinecheck=false]{caption}
\usepackage{subcaption}
\usepackage{multirow}
\usepackage{array}
\usepackage{booktabs}
\usepackage{float}
\usepackage{placeins}
\usepackage{rotating}

\usepackage{amsmath}
\usepackage{bm}
\usepackage{txfonts}
\usepackage{anyfontsize}

\usepackage{enumitem}
\usepackage{hyperref}
\hypersetup{
    colorlinks,
    linkcolor=black,
    citecolor=blue,
    urlcolor=blue
}

\begin{document}

   \title{CHEMOUT: CHEMical complexity in star-forming regions of the OUTer Galaxy}

   \subtitle{V. Chemical composition gradients as a function of the galactocentric radius}

   \author{D. Gigli
          \inst{1, 2}
          \and
          F. Fontani \inst{1, 3, 4}
          \and
          L. Colzi \inst{5}
          \and
          G. Vermari\"en \inst{6}
          \and
          S. Viti \inst{6, 7, 8}
          \and
          V. M. Rivilla \inst{5}
          \and
          A. Sánchez-Monge \inst{9, 10}
          }
    \institute{INAF - Osservatorio Astrofisico di Arcetri, Largo E. Fermi 5, I-50125, Florence, Italy\\
    \email{diegogigli.dg@gmail.com}
    \and
    Dipartimento di Fisica e Astronomia, Università di Firenze, Via G. Sansone 1, 50019 Sesto Fiorentino, Firenze (Italy)
    \and
    Max-Planck-Institut für extraterrestrische Physik, Giessenbachstra\ss{}e 1, 85748 Garching bei M\"unchen, Germany 
    \and
    LUX, Observatoire de Paris, PSL Research University, CNRS, Sorbonne Université, F-92190 Meudon, France 
    \and
    Centro de Astrobiología (CAB), CSIC-INTA, Ctra. de Ajalvir Km. 4, 28850, Torrejón de Ardoz, Madrid, Spain
    \and
    Leiden Observatory, Leiden University, PO Box 9513, 2300 RA Leiden, The Netherlands
    \and
    Transdisciplinary Research Area (TRA) ‘Matter’/Argelander-Institut f\"ur Astronomie, University of Bonn, Bonn, Germany
    \and
    Department of Physics and Astronomy, University College London, Gower Street, London, UK
    \and
    Institut de Ciències de l’Espai (ICE), CSIC, Campus UAB, Carrer de Can Magrans s/n, 08193 Bellaterra (Barcelona), Spain
    \and
    Institut d’Estudis Espacials de Catalunya (IEEC), 08860, Castelldefels (Barcelona), Spain}
   \date{Received 15 June 2025 / Accepted 16 September 2025}

 
  \abstract
   {The outer Galaxy is characterized by lower metallicity compared to regions near the Sun, which suggests that the formation and survival of molecules in star-forming regions within the inner and outer Galaxy are likely to be different.}
   {To understand how chemistry evolves across the Milky Way, deriving molecular abundances in star-forming regions in the outer Galaxy is essential to refining chemical models designed for environments with subsolar metallicity.}
   {We analyzed IRAM 30 m observations in several spectral windows at 3 and 2~mm, toward a sample of 35 sources located at galactocentric distances of $\sim 9-24$~kpc in the context of the project CHEMical complexity in star-forming regions of the outer Galaxy (CHEMOUT).}
   {We focused on the species that have the highest detection rate (i.e., HCN, HCO$^+$, \textit{c}-C$_3$H$_2$, H$^{13}$CO$^+$, HCO, and SO), and searched for possible trends in column densities, abundances, and line widths with the galactocentric distance. We also updated the abundances for H$_2$CO and CH$_3$OH, presented in a previous work, using H$_2$ column densities from new NIKA2 dust continuum maps.
   The fractional abundances with respect to H$_2$ of most of the species (i.e., HCN, HCO$^+$, \textit{c}-C$_3$H$_2$, HCO, H$_2$CO, and CH$_3$OH) scale at most as the elemental fractional abundance of carbon ([C/H]) up to the investigated galactocentric distance of $\sim24~$kpc.
   For the abundances of SO, we find a steeper gradient than that of sulfur elemental abundance ([S/H]). In contrast, the abundances of H$^{13}$CO$^+$ exhibit a shallower gradient relative to that of [$^{13}$C/H].
   Interestingly, we find that gas turbulence, as derived from the full width at half maximum of the lines, decreases with galactocentric distance for all the species investigated, suggesting a more quiescent environment in the outer Galaxy.}
   {These results suggest that, in the outer Galaxy, the efficiency in the formation of most of the molecules studied, scaling with the availability of the parent element, is at least as high as in the local Galaxy, or perhaps even higher (e.g., for H$^{13}$CO$^+$). Yet, other species, such as SO, are characterized by a lower formation efficiency. These outcomes have important implications for the chemistry occurring in the outermost star-forming regions of the Galaxy and can help to constrain models adapted to lower metallicity environments.}


   \maketitle

\section{Introduction}
    \label{sect:1}
    The outer Galaxy (OG), namely the portion of the Galactic disk beyond the solar circle, is located at galactocentric distances ($R_\text{GC}$) approximately between 9 and 24~kpc, and it shows chemical properties significantly different from those of the inner Galaxy. 
    Several chemical galactocentric gradient studies have been conducted, mapping the differences in elemental composition between the inner and outer Milky Way (e.g., C, N, O, Fe), using various tracers (i.e., H\,\textsc{\small II} regions, classical Cepheids, young OB stars, and planetary nebulae, among others). 
    These works, focused on observational trends \citep[e.g.,][]{luck2018cepheid,bragancca2019radial,mendez2022gradients} and chemodynamical studies \citep{jacobson2016gaia,stanghellini2018galactic,palla2020chemical,santos2021ambre}, indicate that the overall metallicity in the OG is lower than the solar one \citep[e.g., a factor of four lower at $R_\text{GC} = 19~$kpc;][]{shimonishi2021detection}. 
    The elemental abundances of carbon, oxygen, and nitrogen - the three most abundant elements in the Universe after hydrogen and helium - decrease as a function of $R_\text{GC}$ \citep[e.g.,][]{esteban2017radial,arellano2020galactic}, similarly to other heavier elements, following the radial metallicity gradient.
    This observational result is linked to the traditional view of the OG as being poorer in molecules than the solar neighborhoods and the inner Galaxy.
    Moreover, based on the aforementioned scarcity of elements crucial for the formation of complex molecules and (rocky) planets, chemical evolution models excluded most of the OG from the so-called Galactic habitable zone (GHZ); that is, the region in the Milky Way where the probability for the development of habitable planets is highest:
    the optimal region is predicted to be an annulus of the Galactic disk from $\sim7~$kpc extended up to $R_\text{GC}\sim9~$kpc, with maximum habitability at $\sim 8$~kpc \citep{spitoni2014galactic,spitoni2017evolution}. 
    This led to the consideration of the OG not being suitable for the formation of planetary systems in which Earth-like planets could be born and might be capable of sustaining life \citep{prantzos2008galactic, ramirez2010possible}.
    
    Recent observational results challenged this prediction.
    First, the occurrence of Earth-like planets does not seem to depend on metallicity \citep[e.g.,][]{maliuk2020spatial,mulders2018exoplanet}.
    Second, the low rate of disruptive events in the OG suggests that it might be more favorable than the inner Galaxy for preserving life on potentially habitable planets \citep[e.g.,][]{piran2014possible,vukotic2016grandeur}.
    Third, the detection of several complex organic molecules (COMs; e.g., CH$_3$OH, CH$_3$CCH, CH$_3$OCHO, CH$_3$OCH$_3$) of prebiotic interest in low-metallicity environments, such as star-forming regions in the OG \citep[][]{shimonishi2021detection,bernal2021methanol,fontani2022chemout,fontani2024chemout,koelemay2025no} and the Magellanic Clouds \citep[][]{sewilo2018detection,sewilo2022alma,shimonishi2023detection, golshan2024high}. 
    Overall, these results suggest that environments prone to the presence of molecular precursors of life and Earth-like planets might be widespread in the Milky Way, including the OG.
    
    The project CHEMical complexity in the outer Galaxy (CHEMOUT; \citeay{fontani2022chemout}, \citeyear{fontani2022chemoutII}, \citeay{colzi2022chemout}, \citeay{fontani2024chemout}), performed with the IRAM 30 m single-dish telescope, is aimed at deriving the molecular composition in 35 star-forming regions of the OG ($R_\text{GC}\sim 9 - 24~$kpc).  
    The main results obtained in the previous works of the CHEMOUT project are (1) both the molecular detection rates and the molecular abundances of some organic species --namely HCO, H$_2$CO, and CH$_3$OH-- do not vary significantly with $R_\text{GC}$ (\citeay{fontani2022chemout}, \citeyear{fontani2022chemoutII});
    (2) the $^{14}$N/$^{15}$N ratio as a function of $R_\text{GC}$ is consistent with chemical-evolution models in which $^{15}$N comes from novae \citep[][]{colzi2022chemout};
    (3) the high-angular-resolution maps of the target with the highest $R_\text{GC}$ (WB89-670 at $R_\text{GC}\sim 23.4$~kpc) reveal chemical differentiation at core scales and, above all, a [C/H] abundance higher than expected at such a high galactocentric distance \citep[][]{fontani2024chemout}.
 
    The results obtained so far indicate a lack of decrease in the inspected molecular abundances, which, however, was limited to three species (HCO, H$_2$CO, and CH$_3$OH) and only 15 sources.
    In this paper, we analyze the molecules detected in more than $60\%$ of the CHEMOUT sample according to \cite{fontani2022chemout}, namely HCN, HCO$^+$, \textit{c}-C$_3$H$_2$, H$^{13}$CO$^+$, HCO, and SO, and revise the molecular abundances of H$_2$CO and CH$_3$OH from \cite{fontani2022chemoutII}, using H$_2$ column densities from the new NIKA2 dust continuum maps. 
    In Sect.~\ref{sect:2}, we describe the observational data used in this paper. 
    In Sect.~\ref{sect:3}, we describe the methods implemented in the data analysis. The results are presented and discussed in Sect.~\ref{sect:4} and \ref{sect:5}. 
    The conclusions and the outlooks are summarized in  Sect.~\ref{sect:6}.
    
\section{Observations}
    \label{sect:2}
    \subsection{IRAM 30 m observations}
    This work is based on the observations done for the CHEMOUT project, described in Papers I and II (\citeay{fontani2022chemout}, \citeyear{fontani2022chemoutII}), performed with the Institut de RadioAstronomie Millimétrique (IRAM) 30 m telescope. 
    The sample of sources analyzed in this paper is composed of the 35 star-forming regions, presented in Table~1 of \citet{fontani2022chemout}.
    In some spectra, two velocity features are identified, and the sources listed in Table~\ref{tab:sources} in this paper include both velocity components. 
    In the analysis conducted, we used spectral windows at 3 and 2 mm, collected with the EMIR receiver and described in Table~\ref{tab:setup}. 
    The spectral windows of the project 004-18 (see Table~\ref{tab:setup}) have been optimized to observe, at 3~mm, the $J=1-0$ transitions of $^{13}$C and $^{15}$N isotopologues of HCN and HNC and, at 2~mm, the $J=2-1$ line of DNC. The spectral windows observed in the project 042-21, at 3 and 2~mm, have been designed to observe several methanol and formaldehyde lines. 
    For further details (e.g., pointing and focus, and weather conditions), we refer the reader to \citeauthor{fontani2022chemout} (\citeyear{fontani2022chemout}, \citeyear{fontani2022chemoutII}).
    
    \subsection{NIKA2 bolometer observations}
    \label{obs:NIKA2}    
    Millimeter continuum emission of 31 out of the 35 CHEMOUT targets was observed with the New IRAM KID Arrays 2 \citep[NIKA2;][]{perotto2020} bolometer of the IRAM 30 m telescope in Pico Veleta, Spain.
    Observations were taken during the winter 2021 and 2023 pool seasons (February). 
    The observations cover an almost circular region with a radius of $\sim 9$ arcmin, centered at the positions where the IRAM 30 m spectra have been observed (Table~\ref{tab:sources}).
    The dual-band capability of NIKA2 allowed us to simultaneously observe the dust's thermal continuum emission at 2.0 and 1.2~mm (150~GHz and 260~GHz, respectively). 
    The sensitivity achieved was $\sim 1.5$ mJy/beam at 1.2~mm and 0.5 mJy/beam at 2.0~mm.
    A more complete description of the observations and a detailed analysis of the maps will be presented in a forthcoming paper of the CHEMOUT series (Fontani et al. in prep). 
    Here, we have used the maps only to derive the H$_2$ column densities (Sect.~\ref{sect:continuummethods}) to compute the molecular fractional abundances with respect to H$_2$ (Sect.~\ref{abundances}).
    \begin{table}[H]
        \centering
        \setlength{\tabcolsep}{2.5pt}
        \caption{Observational set ups of CHEMOUT used in this paper.}
        \begin{tabular}{lccccc}
            \hline
            \hline
            \noalign{\smallskip}
            Spectral windows  & Beam size & $V_\text{res}$ & $B{_\text{eff}}$ & $F_\text{eff}$  & $T_\text{sys}$\\
            (GHz) & ($''$) & (km s$^{-1}$) & & & (K)\\
            \noalign{\smallskip}
            \hline
            \noalign{\smallskip}
            85.310-87.130\tablefootmark{a} & 28 & $\sim$0.16 & 0.80  & 0.95&$\sim120-200$\\
            88.590-90.410\tablefootmark{a} & 27 & $\sim$0.16 & 0.80 & 0.95&$\sim120-200$\\
            90.400-98.180\tablefootmark{b} & 25 & $\sim$0.6 & 0.80 & 0.95&$\sim100-150$\\
            140.720-148.500\tablefootmark{b} & 17 & $\sim$0.4 & 0.73 & 0.93&$\sim150-300$\\
            148.470-150.290\tablefootmark{a} & 15 & $\sim$0.096 & 0.73 & 0.93& $>300$\\
            151.750-153.570\tablefootmark{a} & 15 & $\sim$0.096 & 0.73 & 0.93& $>300$\\
            \noalign{\smallskip}
            \hline
        \end{tabular}
        \tablefoot{
             The observational parameters presented include the frequency ranges observed (at 3 and 2~mm), velocity resolution ($V_\text{res}$) of the spectra, main beam efficiency ($B{_\text{eff}}$), forward efficiency ($F_\text{eff}$), and system temperature ($T_\text{sys}$).\\
            \tablefoottext{a}{Project 004-18}
            \tablefoottext{b}{Project 042-21}
        }
        \label{tab:setup}
    \end{table}
    
\section{Analysis methods}
    \label{sect:3}
    \subsection{Line fitting}
        \label{methods:fit}
        The spectra were fit using the MAdrid Data CUBe Analysis \citep[\textsc{MADCUBA}\footnote{\textsc{MADCUBA} is a software developed at the Madrid Center of Astrobiology (CAB, CSIC-INTA) that enables the visualization and analysis of single spectra and data cubes: \url{https://cab.inta-csic.es/madcuba/}};][]{martin2019spectral} software. The transitions were identified using the Spectral Line Identification and LTE Modelling (SLIM) tool of \textsc{MADCUBA}, which makes use of the Jet Propulsion Laboratory \citep[JPL;][]{pickett1998submillimeter} and Cologne Database for Molecular Spectroscopy \citep[CDMS;][]{muller2001cologne} catalogs. 
        The lines were fit using the AUTOFIT function of SLIM. 
        This function produces the Gaussian synthetic spectrum that best matches the observed spectrum, assuming local thermodynamic equilibrium (LTE) conditions and the same fitting input parameters for all transitions of a given molecular species, which are total column density, $N_\text{tot}$; excitation temperature, $T_\text{ex}$; radial systemic velocity of the source, $V_\text{LSR}$; full width at half-maximum of the line, FWHM; and angular size of the emission, $\theta_\text{S}$. 
        
        For the source size, we assumed that the molecular emission fills the telescope beam, except for methanol and formaldehyde, where an estimation of the emission size was made in \cite{fontani2022chemoutII} and updated in this paper as discussed in Sect.~\ref{sect:updates}. Such an assumption is justified by the spatial extent of the emission of the dust (see Figs.~\ref{fig:NIKA2-1},~\ref{fig:NIKA2-2},~\ref{fig:NIKA2-3}), as well as by the updated H$_2$CO and CH$_3$OH emission sizes shown in Table~\ref{tab:abund2}.
        The excitation temperature, whenever an estimate was not possible through the fit, has been fixed to the one shown in Table~\ref{tab:sources}, which was estimated from the methanol lines for the 15 targets observed in \cite{fontani2022chemoutII}. For the remaining sources, we adopted an average $T_\text{ex}$ of 10.7~K, derived from those sources for which methanol transitions were available. This approximation was necessary because it was not possible to estimate the temperature individually for each source based on the observed transitions. To evaluate the impact of this assumption, we computed $N_\text{tot}$ in the range of excitation temperatures derived from methanol ($7 \leq T_\text{ex} \leq 15~K$; see Table~\ref{tab:sources}). 
        In this $T_\text{ex}$ range, the densities vary from the values shown in Tables~\ref{tab:abund1} and \ref{tab:abund2} by a factor of two at most for the species studied (lowest for HCN, HCO$^+$, and H$^{13}$CO$^+$; highest for SO). 
        
        As a detection criterion, we considered a line as detected if the signal-to-noise ratio (S/N) is greater than three \citep[an estimation of the rms noise is given in][]{fontani2022chemout}, while the detection is considered as tentative for $2.5\leq\text{S/N}\leq3$.
        If a line is undetected (e.g., S/N~$\leq 2.5$), \textsc{MADCUBA} provides a $3\sigma$ upper limit estimate to the line's integrated intensity from Eq. (29) of \citet{martin2019spectral}.
        The upper limit on the integrated intensity is converted into a column-density upper limit using the SLIM parameters $T_\text{ex}$ and $\theta_\text{S}$. For these estimates, the average FWHM for each species across the sample was adopted (Sect.~\ref{dis:widths}). This approach assumes that different molecules trace gases with different physical, kinematic, and excitation conditions.

        The total column densities derived from transitions observed at frequencies corresponding to beam sizes different from 28$''$ (i.e., the largest beam size among the IRAM 30 m telescope spectral windows) have been rescaled by a factor of $(\frac{\theta_\text{beam}}{28''})^2$. For H$_2$CO and CH$_3$OH, the column densities were estimated assuming a specific angular emission size (see Table~\ref{tab:abund2}, and Sect.~\ref{sect:updates}). In cases where the emission size exceeds 28$''$, no correction was applied, under the assumption of uniform emission within the beam. Conversely, when the source size is smaller than 28$''$, the same scaling factor was applied to account for beam dilution effects.
        The estimation of the molecular fractional abundances has been performed using
        \begin{equation}
            X_\text{mol} = \frac{N_\text{tot}}{N_{\text{H}_2}},
            \label{eq:abundances}
        \end{equation}
         where $N_\text{tot}$ is the total molecular column density, and $N_{\text{H}_2}$ is the H$_2$ column density (see Sect.~\ref{sect:continuummethods}).
    \subsection{Continuum analysis}
        \label{sect:continuummethods}
        For this paper, the analysis of these maps is limited to the extraction of the flux density from an angular region of $28~''$ centered on the pointing centers of the molecular spectroscopic data. The dimension of the extracted region has been decided to match the larger beam size of IRAM 30 m in the studied spectral windows (Table~\ref{tab:setup}).
        A complete description and a deeper analysis of the NIKA2 observations will be given in a forthcoming paper (Fontani et al. in prep.).

        The column densities of H$_2$, $N_{\text H_2}$ were derived from the dust-continuum-emission maps.
        In the scales (and densities) traced by our observations, the thermal dust emission is expected to be optically thin. 
        Under this assumption, we can use the following equation to derive $N_{\text H_2}$ from the continuum flux density \cite[e.g.,][]{battersby2011characterizing}:
                \begin{equation}
                    N_{\text H_2}=\frac{\gamma F_\nu}{\Omega_\text{S} \kappa_\nu B_\nu(T_\text{d})\mu_{\text{H}_2} m_\text{H}}\;,
                \end{equation}
        where $\gamma$ is the gas-to-dust ratio, $F_\nu$ is the continuum flux density at frequency $\nu$, $\Omega_\text{S}$ is the source solid angle, $\kappa_\nu$ is the dust-mass opacity, $T_\text{d}$ is the dust temperature, and $\mu_{\text{H}_2}$ is the mean molecular weight for molecular hydrogen that is assumed to be 2.8 \citep{kauffmann2008mambo}. 
        The H$_2$ column density was estimated with both a constant gas-to-dust ratio ($\gamma=100$) and a gas-to-dust ratio increasing with distance from the Galactic center from \cite{giannetti2017galactocentric} ($\gamma \sim 3000$ at R$_\text{GC} = 23.4$~kpc). 
        $T_\text{d}$ was derived under the assumption of thermal coupling between gas and dust, adopting the excitation temperatures obtained from the molecular line analysis of CH$_3$OH (Table~\ref{tab:sources}) or, for those sources where methanol is not detected, adopting 10.7~K (the average $T_\text{ex}$ from CH$_3$OH).
        The dust-mass opacity was derived at the observed frequencies from the relation $\kappa_{\nu}/\kappa_{\nu_0}=(\nu/\nu_0)^{\beta}$, assuming $\kappa_{\nu_0}=0.899$ at 
        $\nu_0 = 230~$GHz \citep{ossenkopf1994dust}.
        The spectral index, $\beta,$ was estimated, for each source, from the ratio between the flux densities at the two frequencies observed (150 and 260~GHz) with NIKA2 (Appendix~\ref{NIKA2}).
        
\section{Results}
    \label{sect:4}
    \subsection{Detection summary}
    \label{sect:summary}
    A comprehensive molecular identification was performed over the whole sample to identify the species detected in the greatest number of targets.
    This allowed us to analyze abundance gradients as a function of galactocentric distance with the highest possible statistics. 
    Over 30 molecular species have been detected, and the most frequently detected ones are  HCN (39/39), HCO$^+$ (39/39), \textit{c}-C$_3$H$_2$ (38/39), H$^{13}$CO$^+$ (30/39), HCO (29/39), and SO (25/39). The detection statistics over these species are already listed in \cite{fontani2022chemout}. We also included H$_2$CO (18/18) and CH$_3$OH (17/18) among the species considered. The analysis of the HCO, H$_2$CO, and CH$_3$OH gradients was conducted by \cite{fontani2022chemoutII}; however, here, we present updated trends based both on the revised fits performed with \textsc{MADCUBA} (Sect.~\ref{sect:updates}) and on our new H$_2$ column-density estimates from NIKA2 observations. The spectroscopic parameters of the transitions detected in these molecules are displayed in Table~\ref{tab:lines}. In Appendix~\ref{app:parameters}, the best-fit parameters obtained from the analysis with \textsc{MADCUBA}, and the spectra of the fit transitions for HCN, H$^{13}$CO$^+$, and SO are shown in detail. The spectra of \textit{c}-C$_3$H$_2$ and HCO$^+$ are shown in \cite{fontani2022chemout}, and the ones of HCO, H$_2$CO, and CH$_3$OH in \cite{fontani2022chemoutII}. The results regarding the molecular column densities (e.g., their galactocentric gradients) are provided in Appendix~\ref{app:densities}. 
    
    In Table~\ref{tab:sources}, the $N_{\text{H}_2}$ values calculated assuming a gas-to-dust ratio from \citet{giannetti2017galactocentric} are presented. These values are used hereafter for the analysis. All the data, results, and discussions concerning fractional abundances with respect to H$_2$ are based on these estimates, rather than on those obtained assuming a fixed $\gamma=100$. The NIKA2 dust continuum maps and the galactocentric gradients for H$_2$ column densities are shown in Appendix~\ref{NIKA2}.
    
    Six sources (19383+2711, 19571+3113, WB89-006, WB89-060, WB89-080, and WB89-380) show multiple velocity features also reported in \cite{fontani2022chemout}. 
    Four of them (19383+2711, 19571+3113, WB89-006, and WB89-380) exhibit double-peak emission in most of the observed lines.
    Because the peak velocities of these two features are consistent in the different transitions in which they are detected, they probably arise from two clumps within the telescope beam that have an angular separation smaller than the beam itself. 
    This is confirmed by the NIKA2 maps (Figs.~\ref{fig:NIKA2-1} - \ref{fig:NIKA2-3}), for 19383+2711 and 19571+3113, showing two different emission components in the beam of the IRAM 30 m telescope. 
    WB89-006 and WB89-380 do not show a clear double peak in the continuum maps, but the two clumps could be non-resolved with NIKA2 or aligned along the line of sight. 
    In principle, each of these clumps may have distinct physical and chemical properties and will therefore be analyzed separately in the following sections. 
    For these, the peak velocities were estimated using the $V_\text{LSR}$ of the \textit{c}-C$_3$H$_2$ $J_{K_a,K_c} = 2_{1,2} - 1_{0,1}$ line, and we assumed the same Galactic longitude, $\ell$, for both clumps, to estimate their galactocentric radius, $R_\text{GC,}$ and heliocentric distance, $d$. 
    The remaining two sources (WB89-060 and WB89-080) are characterized by a double peak, but only in some more intense and potentially optically thick lines (e.g., HCN and HCO$^+$) and do not show coherence in the peak velocities. Consequently, in these cases, the double-peaked line shapes are more likely due to high-optical-depth effects, such as (self-)absorption, rather than multiple velocity features. 
    These have thus been analyzed as single sources, and the absorbed lines discarded from the analysis. 
    Moreover, some transitions showed non-Gaussian, high-velocity wings. 
    We decided to limit our analysis to the narrower component of the molecular emissions related to the line emission around the intensity peak, and thus arising from the bulk emission of the cores.
        \begin{table}[H]
            \centering
            \caption{Observed ranges of column densities, fractional abundances, and line widths of the studied species.}
            \begin{tabular}{lccc}
                \hline
                \hline
                \noalign{\smallskip}
                Species & $N_\text{tot}$ & $X_\text{mol}$ & FWHM \\
                 & (cm$^{-2}$) & & (km s$^{-1}$) \\
                \noalign{\smallskip}
                \hline
                \noalign{\smallskip}
                HCN & $(0.33 - 24) \times 10^{12}$ & $(0.84 - 17) \times 10^{-11}$ & $0.6 - 3.8$ \\
                HCO$^+$ & $(0.31 - 8.8)\times 10^{12}$ & $(0.34 - 6.8)\times 10^{-11}$ & $0.9 - 4.3$ \\
                \textit{c}-C$_3$H$_2$ & $(0.31 - 3.5) \times 10^{12}$ & ($0.34 - 10) \times 10^{-11}$ & $0.7 - 4.7$\\
                H$^{13}$CO$^+$ & $(0.31 - 4.8) \times 10^{11}$ & $(0.43 - 10) \times 10^{-12}$ & $0.6 - 4.9$ \\
                HCO & $(0.58 - 7.6) \times 10^{12}$ & $(0.57 - 23) \times 10^{-11}$ & $0.7 - 4.8$ \\
                SO & $(0.24 - 5.2) \times 10^{13}$ & $(0.21 - 5.2)\times 10^{-10}$ & $0.6 - 4.1$ \\ 
                H$_2$CO & $(0.15-6.5)\times10^{13}$ & $(0.26 - 3.7)\times10^{-10}$ & $1.6-4.0$\\
                CH$_3$OH & $(0.24-11)\times10^{13}$ & $(0.15-5.6)\times10^{-10}$ & $1.0-4.0$\\
                \noalign{\smallskip}
                \hline
            \end{tabular}
            \label{tab:ranges}
        \end{table}
        
        \begin{table*}
            \centering
            \caption{Rest frequencies and other spectroscopic parameters of the studied and detected molecular transitions.}
            \begin{tabular}{lcccc}
                \hline
                \hline
                \noalign{\smallskip}
                Molecule & Rest frequency  & Quantum numbers & $E_\text{u}$ & log($A_\text{ij}$/s) \\
                 & (GHz) & & (K) & \\
                \noalign{\smallskip}
                \hline
                \noalign{\smallskip}
                HCN & 88.63042 & $J = 1-0, F = 1-1$ & 4.3 & $-4.6184$ \\
                &88.63185 & $J = 1-0, F = 2-1$ & 4.3 & $-4.6185$ \\
                &88.63394 &$ J = 1-0, F = 0-1$ & 4.3 & $-4.6184$ \\
                \noalign{\smallskip}
                \hline
                \noalign{\smallskip}
                HCO$^+$ & 89.18852 & $J = 1-0$ & 4.3 & $-4.3781$ \\
                \noalign{\smallskip}
                \hline
                \noalign{\smallskip}
                \textit{c}-C$_3$H$_2$ & 85.33889 & ${J}_{{K}_a, {K}_c} = 2_{1,2} - 1_{0,1}$ & 6.4 & $-4.6341$\\
                & 145.08959 & ${J}_{{K}_a, {K}_c} = 3_{1,2} - 2_{2,1}$ & 16.0 & $-4.1696$ \\
                \noalign{\smallskip}
                \hline
                \noalign{\smallskip}
                H$^{13}$CO$^+$ & 86.75429 & $J = 1-0 $& 4.2 & $-4.4142$ \\
                \noalign{\smallskip}
                \hline
                \noalign{\smallskip}
                HCO & 86.67076 & ${N}_{K_a,K_c} = 1_{0,1} - 0_{0,0}, J = 3/2-1/2, F = 2-1$ & 4.2 & $-5.3289$ \\
                & 86.70836 & ${N}_{K_a,K_c} = 1_{0,1} - 0_{0,0}, J = 3/2-1/2, F = 1-0$ & 4.2 & $-5.3377$ \\
                & 86.77746 & ${N}_{K_a,K_c} = 1_{0,1} - 0_{0,0}, J = 1/2-1/2, F = 1-1$ & 4.2 & $-5.3366$ \\
                & 86.80578 & ${N}_{K_a,K_c} = 1_{0,1} - 0_{0,0}, J = 1/2-1/2, F = 0-1$ & 4.2 & $-5.3268$ \\
                \noalign{\smallskip}
                \hline
                \noalign{\smallskip}
                SO & 86.09395 & $N = 2-1, J = 2-1$ & 19.3 & $-5.2799$ \\
                \noalign{\smallskip}
                \hline
                \noalign{\smallskip}
                H$_2$CO & 140.839502 & ${J}_{{K}_a, {K}_c} = 2_{1,2} - 1_{1,1}$ & 21.9 & $-4.2754$\\
                H$_2$CO & 145.602949 & ${J}_{{K}_a, {K}_c} = 2_{0,2} - 1_{0,1}$ & 10.5 & $-4.1072$      \\
                \noalign{\smallskip}
                \hline
                \noalign{\smallskip}
                CH$_3$OH & 95.91431 & $2(1,2)-1(1,1)$ A$^+$ & 21.4 & $-5.6031$\\
                CH$_3$OH & 96.739358 & $2(1,2)-1(1,1)$ E$_2$ & 12.5 & $-5.5923$\\
                CH$_3$OH & 96.741371 & $2(0,2)-1(0,1)$ A$^+$ & 7.0 & $-5.4675$\\
                CH$_3$OH & 96.744545 & $2(0,2)-1(0,1)$ E$_1$ & 20.1 & $-5.4676$\\
                CH$_3$OH & 96.755501 & $2(1,1)-1(1,0)$ E$_1$ & 28.0 & $-5.5810$\\
                CH$_3$OH & 97.582798 & $2(1,1)-1(1,0)$ A$^-$ & 21.6 & $-5.5806$\\
                CH$_3$OH & 143.865795 & $3(1,3)-2(1,2)$ A$^+$ & 28.3 & $-4.9712$\\
                CH$_3$OH & 145.093707 & $3(0,3)-2(0,2)$ E$_1$ & 27.1 & $-4.9096$\\
                CH$_3$OH & 145.097370 & $3(1,3)-2(1,2)$ E$_2$ & 19.5 & $-4.9602$\\
                CH$_3$OH & 145.103152 & $3(0,3)-2(0,2)$ A$^+$ & 13.9 & $-4.9093$\\
                CH$_3$OH & 145.126191 & $3(2,2)-2(2,1)$ E$_1$ & 36.2 & $-5.1693$\\
                CH$_3$OH & 145.126386 & $3(2,1)-2(2,0)$ E$_2$ & 39.8 & $-5.1638$\\
                CH$_3$OH & 145.131855 & $3(1,2)-2(1,1)$ E$_1$ & 35.0 & $-4.9490$\\
                CH$_3$OH & 146.368328 & $3(1,2)-2(1,1)$ A$^+$ & 28.6 & $-4.9488$\\
                \noalign{\smallskip}
                \hline
            \end{tabular}
            \label{tab:lines}
        \end{table*}
        \subsubsection{Formaldehyde and methanol updated fits}
            \label{sect:updates}
            The fits for H$_2$CO and CH$_3$OH have been revised due to an incorrect assumption in \cite{fontani2022chemoutII}, which originated from a conversion between $T_\mathrm{A}^*$ and $T_{\rm MB}$. Since the beam efficiency $\eta_\text{eff}$ of the telescope was not explicitly specified in \cite{bernal2021methanol}, the values were taken from Appendix C.3 of the ARO 12m user manual; however, outdated efficiencies were inadvertently used. Updated efficiencies\footnote{From https://aro.as.arizona.edu/?q=beam-efficiencies.} have now been adopted to estimate the $T_\mathrm{MB}$ of these two species. For H$_2$CO at 2~mm we divided $T_\mathrm{A}^*$ by 0.81, and for CH$_3$OH at 3~mm we divided by 0.91. 
            The source sizes for H$_2$CO and CH$_3$OH were re-estimated using Eq.~3 in \cite{fontani2002structure}. Consequently, the MADCUBA fits were recalculated, and the resulting parameters are reported in Appendix~\ref{app:parameters}. The excitation temperatures of CH$_3$OH are listed in Table~\ref{tab:sources}.
        
    \subsection{Fractional abundances}
        \label{abundances}
        The molecular fractional abundances, $X_\text{mol}$, with respect to H$_2$ were computed through Eq.~\ref{eq:abundances}, with the molecular column densities from Tables~\ref{tab:abund1} and \ref{tab:abund2}, and the $N_{\text H_2}$ from Table~\ref{tab:sources}. The values of $X_\text{mol}$ are in ranges presented in Table~\ref{tab:ranges}.
        All the values are shown in Table~\ref{tab:abundances}. These values are estimated for 31 star-forming regions of the sample, i.e., the ones observed in the dust emission, for which we have the H$_2$ column densities. For sources exhibiting a double clump structure (see Sect.~\ref{sect:summary}), the fractional abundances were obtained by summing the molecular column densities of the two components, and the galactocentric distances were determined as the average $R_\text{GC}$ of the individual clumps.  
        \begin{table*}
            \centering
            \caption{Fractional abundance, $X_\text{mol}$, with respect to H$_2$ of HCN, HCO$^+$, \textit{c}-C$_3$H$_2$, H$^{13}$CO$^+$, HCO, SO, H$_2$CO, and CH$_3$OH.}
            \begin{tabular}{lcccccccc}
                \hline
                \hline
                \noalign{\smallskip}
                Source & X(HCN) & X(HCO$^+$) & X(\textit{c}-C$_3$H$_2$) & X(H$^{13}$CO$^+$) & X(HCO) & X(SO) & X(H$_2$CO) & X(CH$_3$OH)\\
                &  $\times 10^{-11}$ & $\times 10^{-11}$ & $\times 10^{-11}$  & $\times 10^{-12}$ &$\times 10^{-11}$ & $\times 10^{-10}$ & $\times 10^{-10}$ & $\times 10^{-10}$\\
                \noalign{\smallskip}
                \hline
                \noalign{\smallskip}
                19383+2711\tablefootmark{a} & 6.7 (1.5) & 3.1 (0.7) & 1.8 (0.4) & 0.9 (0.3) & 3.1 (1.3) & 0.36 (0.13) & 0.49 (0.15) & $\leq$0.6 \\
                19423+2541 & 7.0 (1.7) & 2.1 (0.6) & 1.2 (0.3) & 0.7 (0.2) & 1.4 (0.4) & 0.8 (0.2) & 2.2 (0.7) & 1.6 (0.4) \\
                19489+3030 & 6.7 (1.5) & 4.9 (1.0) & 2.5 (0.6) & 4.6 (0.12) & $\leq$1.0 & 0.5 (0.3) & ... & ... \\
                19571+3113\tablefootmark{a} & 7 (2) & 4.7 (1.1) & 4.0 (1.1) & $\leq$0.9 & 7 (4) & $\leq$0.4 & ... & ... \\
                20243+3853 & 8.3 (1.7) & 4.0 (0.9) & 2.4 (0.5) & 1.3 (0.4) & 4.3 (1.0) & 0.7 (0.2) & ... & ... \\
                WB89-002\tablefootmark{b} & ... & ... & ... & ... & ... & ... & ... & ... \\
                WB89-006\tablefootmark{a} & 3.1 (1.6) & 2.3 (0.6) & 1.8 (0.7) & 2.6 (0.9) & $\leq$1.6 & $\leq$0.5 & 0.44 (0.13) & 2.2 (0.8) \\
                WB89-014 & 4.3 (1.1) & 3.6 (0.9) & 2.5 (0.8) & $\leq$1.3 & $\leq$3 & $\leq$0.8 & ... & ... \\
                WB89-031 & 3.2 (0.7) & 1.7 (0.5) & 2.0 (0.6) & 0.8 (0.4) & 3.2 (1.2) & $\leq$0.4 & ... & ... \\
                WB89-035 & 9 (2) & 4.8 (1.2) & 3.8 (1.1) & 2.2 (0.7) & 3.5 (1.3) & 2.0 (0.6) & 3.7 (1.1) & 3.2 (1.0) \\
                WB89-040 & 17 (5) & 2.4 (1.0) & 10 (3) & 10.2 (0.4) & 23 (8) & 5 (2) & ... & ... \\
                WB89-060 & ... & ... & 0.8 (0.3) & 2.9 (0.7) & $\leq$0.4 & 2.2 (0.6) & ... & ... \\
                WB89-076 & 3.2 (0.8) & 0.9 (0.3) & 1.8 (0.4) & 2.0 (0.5) & 0.6 (0.2) & 1.6 (0.4) & 0.63 (0.17) & 1.6 (0.6) \\
                WB89-080 & 4.9 (1.1) & ... & 1.9 (0.7) & 1.6 (0.6) & 1.4 (0.5) & 1.5 (0.4) & 2.3 (0.7) & 2.7 (0.8) \\
                WB89-083 & 7.4 (1.5) & 6.8 (1.4) & 3.6 (0.8) & 2.3 (0.6) & 5.2 (1.8) & $\leq$0.7 & ... & ... \\
                WB89-152\tablefootmark{b} & ... & ... & ... & ... & ... & ... & ... & ... \\
                WB89-283 & 7.0 (1.8) & 2.7 (0.7) & 2.0 (0.7) & 0.8 (0.3) & 7 (2) & 0.8 (0.3) & 3.2 (1.0) & 3.1 (0.8) \\
                WB89-288 & 1.5 (0.3) & 2.1 (0.4) & 1.0 (0.3) & $\leq$0.5 & 3.4 (1.0) & $\leq$0.3 & ... & ... \\
                WB89-315\tablefootmark{b} & ... & ... & ... & ... & ... & ... & ... & ... \\
                WB89-379 & 3.3 (0.8) & 0.5 (0.2) & 0.49 (0.14) & 0.7 (0.2) & 1.3 (0.4) & 0.3 (0.11) & 0.9 (0.2) & 1.2 (0.4) \\
                WB89-380\tablefootmark{a} & 2.6 (0.7) & 1.4 (0.4) & 0.8 (0.8) & 0.6 (0.2) & $\leq$1.3 & 0.29 (0.10) & 0.26 (0.12) & 0.54 (0.17)\\
                WB89-391 & 3.6 (0.9) & 1.7 (0.4) & 1.9 (0.5) & 0.63 (0.19) & 2.4 (0.6) & 0.9 (0.2) & 0.7 (0.2) & 1.5 (0.4) \\
                WB89-399 & 2.7 (0.6) & 1.8 (0.4) & 1.5 (0.3) & $\leq$0.2 & 2.2 (0.6) & $\leq$0.13 & 0.6 (0.2) & 0.15 (0.07) \\
                WB89-437 & 11 (3) & 4.2 (0.9) & 1.2 (0.3) & 1.6 (0.4) & 1.0 (0.3) & 1.7 (0.4) & 2.6 (0.7) & 5.4 (1.3) \\
                WB89-440 & 2.7 (0.5) & 4.3 (0.8) & 1.7 (0.4) & $\leq$0.4 & 5.9 (1.4) & $\leq$0.2 & ... & ... \\
                WB89-501 & 5.2 (1.2) & 2.3 (0.6) & 1.3 (0.3) & 0.7 (0.2) & 2.6 (0.7) & 0.53 (0.14) & 1.8 (0.5) & 1.6 (0.5) \\
                WB89-529 & 0.84 (0.19) & 1.7 (0.3) & 0.7 (0.2) & $\leq$0.4 & 2.3 (0.6) & $\leq$0.2 & ... & ... \\
                WB89-572 & 1.8 (0.4) & 0.9 (0.2) & 0.34 (0.14) & 0.4 (0.2) & $\leq$0.7 & 0.6 (0.2) & ... & ... \\
                WB89-621 & 6.9 (1.7) & 2.0 (0.5) & 0.49 (0.13) & 1.4 (0.4) & 0.8 (0.3) & 2.9 (0.7) & 2.1 (0.6) & 5.6 (1.5) \\
                WB89-640 & 2.2 (0.4) & 0.34 (0.11) & 0.80 (0.19) & 1.4 (0.3) & 2.6 (0.6) & 0.35 (0.11) & ... & ... \\
                WB89-670 & 1.0 (0.3) & 0.8 (0.2) & 1.3 (0.5) & 1.6 (0.6) & $\leq$0.6 & $\leq$0.18 & ... & ... \\
                WB89-705 & 1.5 (0.4) & 0.9 (0.2) & 1.9 (0.5) & 2.9 (0.8) & 1.0 (0.4) & 0.67 (0.18) & ... & ... \\
                WB89-789 & 1.3 (0.3) & 0.87 (0.19) & 0.35 (0.11) & 0.8 (0.2) & 0.7 (0.2) & 0.21 (0.07) & 0.5 (0.2) & 0.37 (0.11) \\
                WB89-793 & 4.8 (1.2) & 2.1 (0.5) & 1.4 (0.5) & 2.5 (0.7) & $\leq$1.1 & 1.2 (0.4) & ... & ... \\
                WB89-898\tablefootmark{b} & ... & ... & ... & ... & ... & ... & ... & ... \\
                \noalign{\smallskip}
                \hline
            \end{tabular}
            \tablefoot{
            \tablefoottext{a}{Sources characterized by a double clump. The fractional abundances have been estimated by summing the total column densities over both clumps.}
            \tablefoottext{b}{Sources lacking NIKA2 observations, for which no H$_2$ column density could be estimated.}
            } 
            \label{tab:abundances}
        \end{table*}

\begin{figure*}
            \centering
            \includegraphics[width=0.97\linewidth]{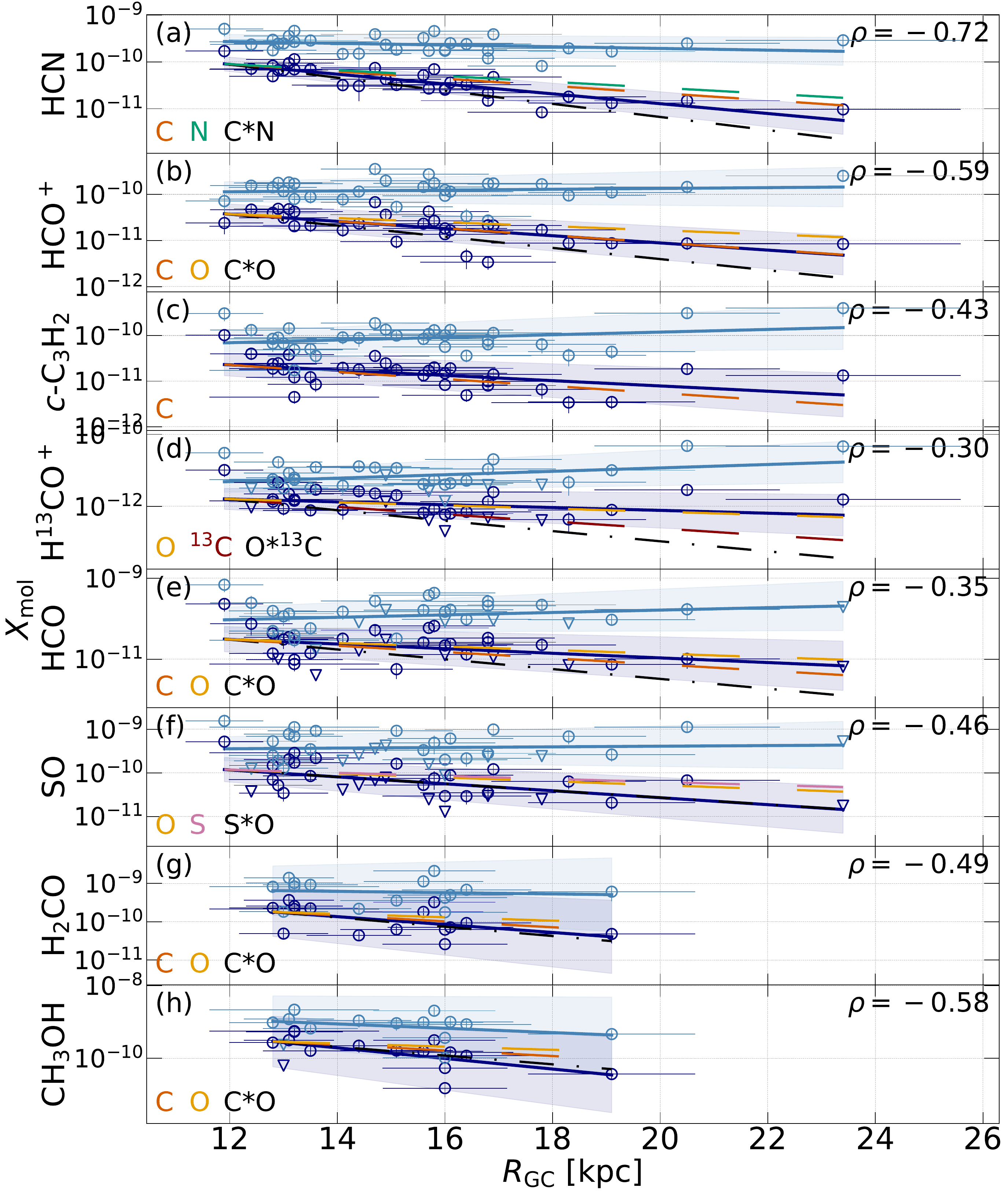}
            \caption{Galactocentric gradients of fractional abundances with respect to H$_2$, $X_\text{mol}$. The plot shows the trends for HCN (a), HCO$^+$(b), \textit{c}-C$_3$H$_2$(c), H$^{13}$CO$^+$(d), HCO (e), SO (f), H$_2$CO (g), and CH$_3$OH (h) as a function of the galactocentric radius ($R_\text{GC}$). The light blue data represent the abundances calculated using a constant gas-to-dust ratio ($\gamma=100$), while the dark blue data illustrate those estimated using a non-constant gas-to-dust ratio \citep[$\gamma = \gamma(R_\text{GC})$ from][]{giannetti2017galactocentric}. For both datasets, the linear regression results are shown as the light blue and dark blue lines, respectively. The $1\sigma$ error bars over the slope of the gradients are displayed for each fit. The upper-limit values are represented with triangles. The gradients of the elemental abundances of carbon (C and $^{13}$C), nitrogen (N), oxygen (O), and sulfur (S), as reported by \citet{mendez2022gradients}, are plotted as dashed lines. The product of the parent elements of the species is represented, in each subplot, by a dash–dotted black line. All elemental trends are plotted with a vertical offset to align with the starting point of the molecular gradients derived using $\gamma(R_\text{GC})$, to facilitate comparison of their slopes with the linear fit of the molecular abundances. In the upper-right side of each subplot, the Pearson correlation coefficient, $\rho$, is shown (estimated only for the abundances estimated using the non-constant gas-to-dust ratio).}
            \label{fig:abundances}
        \end{figure*}

        The plots showing $X_\text{mol}$ against $R_\text{GC}$ for each molecule are shown in Fig.~\ref{fig:abundances}. For comparison, the fractional abundances calculated using a gas-to-dust ratio $\gamma = 100$ are also shown. 
        A linear fit is performed for each plot, and the Pearson correlation coefficient\footnote{This coefficient measures the strength and direction of the linear relationship between two variables, X and Y, and it is calculated as the covariance of X and Y divided by the product of their standard deviations. It ranges from $-1$ to $+1$, where $+1$ indicates perfect positive correlation, 0 indicates no linear correlation, and $-1$ indicates perfect negative correlation.}, $\rho$, is calculated to inspect possible (anti-)correlations.  
        The linear fit applied to the data of fractional abundances, calculated using a gas-to-dust ratio from \cite{giannetti2017galactocentric}, gives: 
        {\fontsize{8.5}{10}\selectfont
            \begin{align}
                [\text{HCN}] &= (-0.105 \pm 0.013)\ \mathrm{dex\ kpc}^{-1} \cdot R_\text{GC} - (8.79 \pm 0.20)\ \mathrm{dex} \label{eq:grad1},\\
                [\text{HCO}^+] &= (-0.078 \pm 0.018)\ \mathrm{dex\ kpc}^{-1} \cdot R_\text{GC} - (9.50 \pm 0.28)\ \mathrm{dex} \label{eq:grad2},\\
                [\textit{c}\text{-C}_3\text{H}_2] &= (-0.058 \pm 0.021)\ \mathrm{dex\ kpc}^{-1} \cdot R_\text{GC} - (9.95 \pm 0.32)\ \mathrm{dex} \label{eq:grad3},\\
                [\text{H}^{13}\text{CO}^+] &= (-0.039 \pm 0.025)\ \mathrm{dex\ kpc}^{-1} \cdot R_\text{GC} - (11.32 \pm 0.38)\ \mathrm{dex} \label{eq:grad4},\\
                [\text{HCO}] &= (-0.057 \pm 0.026)\ \mathrm{dex\ kpc}^{-1} \cdot R_\text{GC} - (9.83 \pm 0.41)\ \mathrm{dex} \label{eq:grad5},\\              
                [\text{SO}] &= (-0.080 \pm 0.023)\ \mathrm{dex\ kpc}^{-1} \cdot R_\text{GC} - (8.98 \pm 0.36)\ \mathrm{dex} \label{eq:grad6} ,\\
                [\text{H}_2\text{CO}] &= (-0.103 \pm 0.050)\ \mathrm{dex\ kpc}^{-1} \cdot R_\text{GC} - (8.42 \pm 0.75)\ \mathrm{dex} \label{eq:grad7},\\
                [\text{CH}_3\text{OH}] &= (-0.146 \pm 0.054)\ \mathrm{dex\ kpc}^{-1} \cdot R_\text{GC} - (7.67 \pm 0.82)\ \mathrm{dex.} \label{eq:grad8}
            \end{align}
        }
        
    \subsection{Line widths}
        \label{sect:width}
        From the fit of the lines, we also obtained the full widths at half maximum (FWHMs).
        The measured FWHMs for the molecules studied are in the ranges shown in Table~\ref{tab:ranges}. 
        The mean values of line width estimated for each molecule have been used in the estimation of the upper limits of the column densities, as an educated guess ($2.1\pm0.6~$km s$^{-1}$ for HCN; $1.8\pm0.9~$km s$^{-1}$ for HCO$^+$; $2.1\pm1.0~$km s$^{-1}$  for \textit{c}-C$_3$H$_2$; $2.1\pm1.0~$km s$^{-1}$ for H$^{13}$CO$^+$; $2.2\pm1.0~$km s$^{-1}$ for HCO; $2.0\pm0.8~$km s$^{-1}$ for SO; $2.3 \pm 0.7~\mathrm{km\ s^{-1}}$ for H$_2$CO; $1.9\pm0.8~\mathrm{km\ s^{-1}}$ for CH$_3$OH).
        
\begin{figure*}
            \centering
            \includegraphics[width=\linewidth]{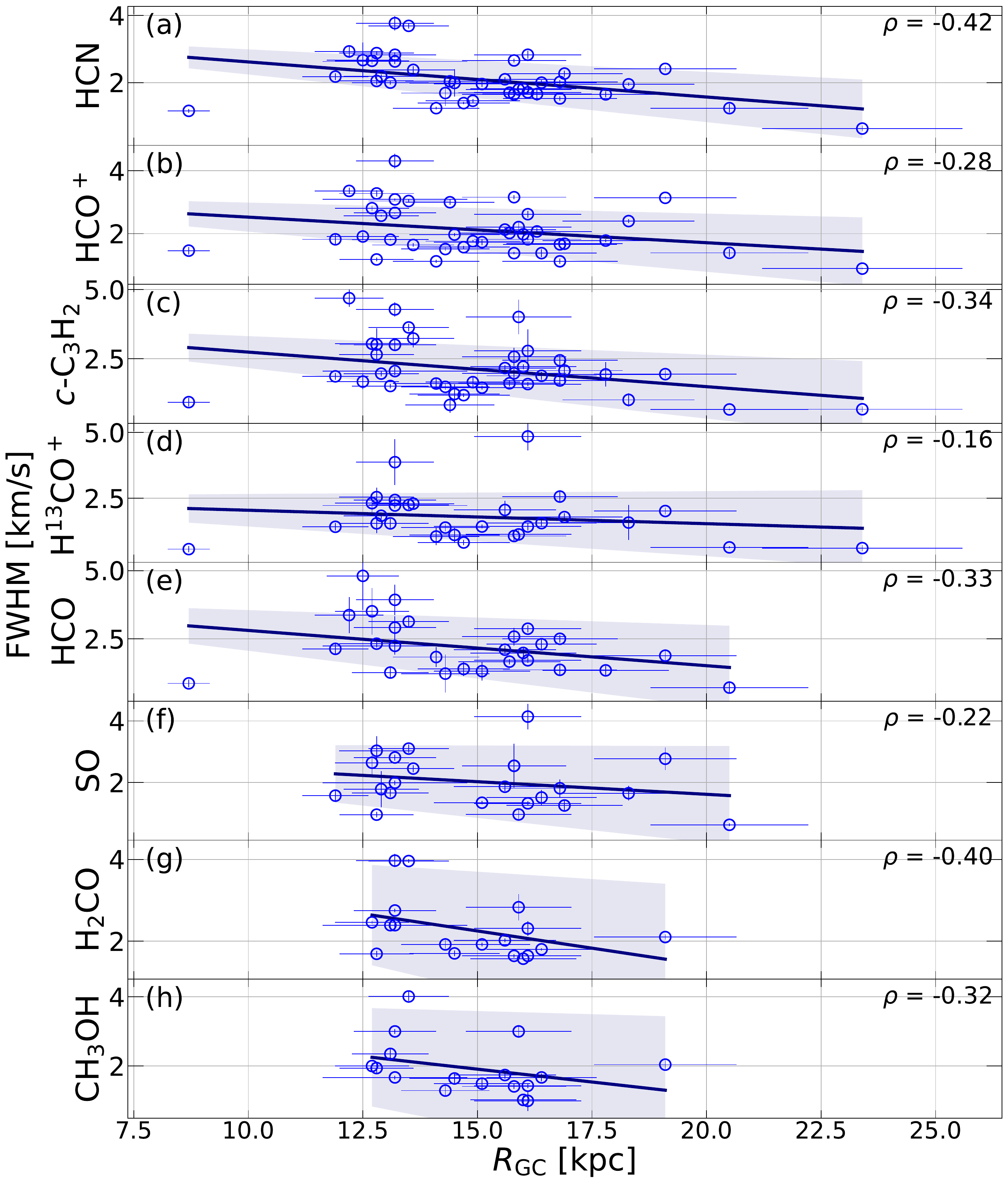}
            \caption{Galactocentric gradients of line widths, FWHM. The plot shows the trends for HCN (a), HCO$^+$(b), \textit{c}-C$_3$H$_2$(c), H$^{13}$CO$^+$(d), HCO (e), SO (f), H$_2$CO (g), and CH$_3$OH (h) as a function of the galactocentric radius ($R_\text{GC}$). The linear fit results are shown as dark blue lines. The $1\sigma$ error bars over the slope of the gradients are displayed for each fit. In the upper right side of each subplot, the Pearson correlation coefficient, $\rho$, is shown.}
            \label{fig:fwhm1}
        \end{figure*}

        The plots of FWHM of HCN, HCO$^+$, \textit{c}-C$_3$H$_2$, H$^{13}$CO$^+$, HCO, SO, H$_2$CO, and CH$_3$OH as a function of the galactocentric distance are shown in Fig.~\ref{fig:fwhm1}. 
        A linear fit is performed, and the Pearson correlation coefficient, $\rho$, is estimated. 
        We find that, for all species, the line width shows an overall negative trend with $R_\text{GC}$, along with a weak negative correlation ($\rho \lesssim -0.2$).

\section{Discussion}
    \label{sect:5}
    \subsection{Galactocentric gradients of fractional abundances}
        \label{dis:abund}
        The trends obtained for the molecular abundance (with respect to H$_2$) were compared with the galactocentric abundance gradients of carbon (both $^{12}$C and $^{13}$C), oxygen, nitrogen, and sulfur. These are constituting elements of the studied species, and the gradients were determined in \cite{mendez2022gradients} through the analysis of the spectra of Galactic H\,\textsc{\small II} regions. The galactocentric gradient of $^{13}$C was estimated by coupling the carbon gradient with the galactocentric trend of the $^{13}\text{C}/^{12}\text{C}$ ratio from \cite{yan2023direct}. Among 8.34~kpc, the galactocentric distance of the Sun, and 24~kpc, which is the distance of the most distant source in our sample, the abundance ratios [C/H], [$^{13}$C/H], [N/H], [O/H], and [S/H] decrease by factors of approximately $\sim16$, $\sim36$, $\sim10$, $\sim5$, and $\sim4$, respectively. Extrapolating the gradients derived in the OG (Eqs.~\ref{eq:grad1}–\ref{eq:grad6}) to the Solar circle, over the same range of $R_\text{GC}$, we find that the abundance of the studied species, i.e., HCN, HCO$^+$, \textit{c}-C$_3$H$_2$, H$^{13}$CO$^+$, HCO, SO, H$_2$CO, and CH$_3$OH decrease by a factor of $\sim44$, $\sim17$, $\sim8$, $\sim4$, $\sim8$, $\sim18$, $\sim41$, and $\sim194.$ 
        
        The decreasing rates of HCN, HCO$^+$, \textit{c}-C$_3$H$_2$, HCO, H$_2$CO, and CH$_3$OH are consistent, within the error bars, with that of [C/H]. 
        This implies that, considering the dispersion and uncertainties in estimating these quantities, the fractional abundances of these carbon-bearing species scale with the elemental abundance of carbon ([C/H]) between 8.34 kpc and 24 kpc. This suggests that the efficiency in forming these molecules, which scales with the availability of the parent element, C, is at least as high as in the local Galaxy.
    
        In contrast, the decreasing rate of SO appears steeper than that of [S/H]. 
        For SO, the higher decreasing rate with respect to [S/H] indicates a lower efficiency of the formation of these molecules in the OG with respect to the local Galaxy. 
        We speculate that a steeper decreasing trend for SO, which is the best shock tracer among those that we analyzed in this work, could indicate that, in the OG, outflows have a reduced influence on the gas composition compared to the local Galaxy.
        This result would be consistent with the decreasing line width with $R_\text{GC}$ (Sect.~\ref{sect:width}).  
        Another shock tracer is CH$_3$OH, which, although its decreasing trend is marginally consistent with that of carbon, exhibits the most pronounced reduction among the species investigated in the OG. The steep decreasing trend of CH$_3$OH could also arise from a less efficient grain-surface chemistry in the OG. 
        The results concerning HCO, H$_2$CO, and CH$_3$OH, obtained with the updated abundances (and distances), confirm the ones of \cite{fontani2022chemoutII}.
        In \cite{fontani2024chemout}, a correlation between the abundances of CH$_3$OH and SO was found, which is confirmed over the entire CHEMOUT sample, although no correlation between the line widths of these two species is observed, hinting that they are not tracing the same gas components.
         
        The decreasing rate of H$^{13}$CO$^+$, although lower than that of [$^{13}$C/H], aligns well with the decreasing rate observed for oxygen. This correlation is not observed for HCO$^+$, whose decreasing rate is consistent with carbon rather than oxygen. This discrepancy could suggest a possible revision of the galactocentric gradient of the $^{13}$C/$^{12}$C ratio, with a shallower trend at high $R_\text{GC}$, in agreement with the Galactic chemical evolution discussed in \cite{colzi2022chemout}.
        Moreover, the detailed analysis of the $^{13}$C/$^{12}$C ratios from the HCO$^+$ and H$^{13}$CO$^+$ lines will be the subject of a forthcoming paper (Colzi et al. in prep.).
        Therefore, the results regarding \textit{c}-C$_3$H$_2$, HCO, and especially H$^{13}$CO$^+$ lead to the puzzling conclusion of observing more carbon-bearing molecules than the available carbon atoms in the OG, their decreasing rates being lower than that of [C/H] (or [$^{13}$C/H]).        
    \subsection{Galactocentric gradients of line widths}
        \label{dis:widths}
        Considering the low temperature (Table~\ref{tab:sources}) of the sources, the thermal broadening of the emission lines can be neglected ($\sim0.1-0.2$~km s$^{-1}$); therefore, the line width is a good tracer of the gas turbulence of the sources.
        The decreasing galactocentric trends observed for the FWHM of the studied molecules suggest that the sources tend to have lower turbulence as the distance from the Galactic center increases.
        Such a result would be in line with an environment becoming more and more quiescent at increasing $R_\text{GC}$.
        The nearest source to the Sun, WB89-002, is the only outlier that deviates from these trends.
        
        This result is also consistent with the study of \cite{lin2025inadequate}, where the velocity dispersion of low-$J$ $^{13}$CO lines has been studied toward molecular clouds in the low-metallicity part of the Galactic disk. A decreasing trend of the line width is observed with $R_\text{GC}$, and the sub-virial dynamics observed in the metal-poor clouds hint at how the gas turbulence is not sufficient to counterbalance the cloud's self-gravity. 
        As stated in \citet{lin2025inadequate}, in low-metallicity environments, the magnetic field may play a more dominant role, replacing gas turbulence in sustaining the clouds.
    \subsection{Galactocentric gradients of molecular ratios}
        \label{dis:ratios}
        Studies on the formation pathways of HCO$^+$, H$^{13}$CO$^+$ \citep{klemperer1970carrier, millar1991gas}, HCO, H$_2$CO, and CH$_3$OH \citep{watanabe2002efficient, fuchs2009hydrogenation, minissale2016dust, rivilla2019first, jimenez2025modelling} have shown that their formation is linked to the presence of CO. In contrast, \textit{c}-C$_3$H$_2$ is formed from "free" carbon (not bound in CO) \citep{thaddeus1985laboratory,sipila2016understanding}. Moreover, it is known that [C/O] decreases with distance from the Galactic center \citep{mendez2022gradients}. Therefore, we expect to have less free carbon in the OG to produce \textit{c}-C$_3$H$_2$, because most of it would be locked in CO. This would lead to a prediction of an increasing trend for the CO/\textit{c}-C$_3$H$_2$ ratio.
    
\begin{figure*}
            \centering
            \includegraphics[width=\linewidth]{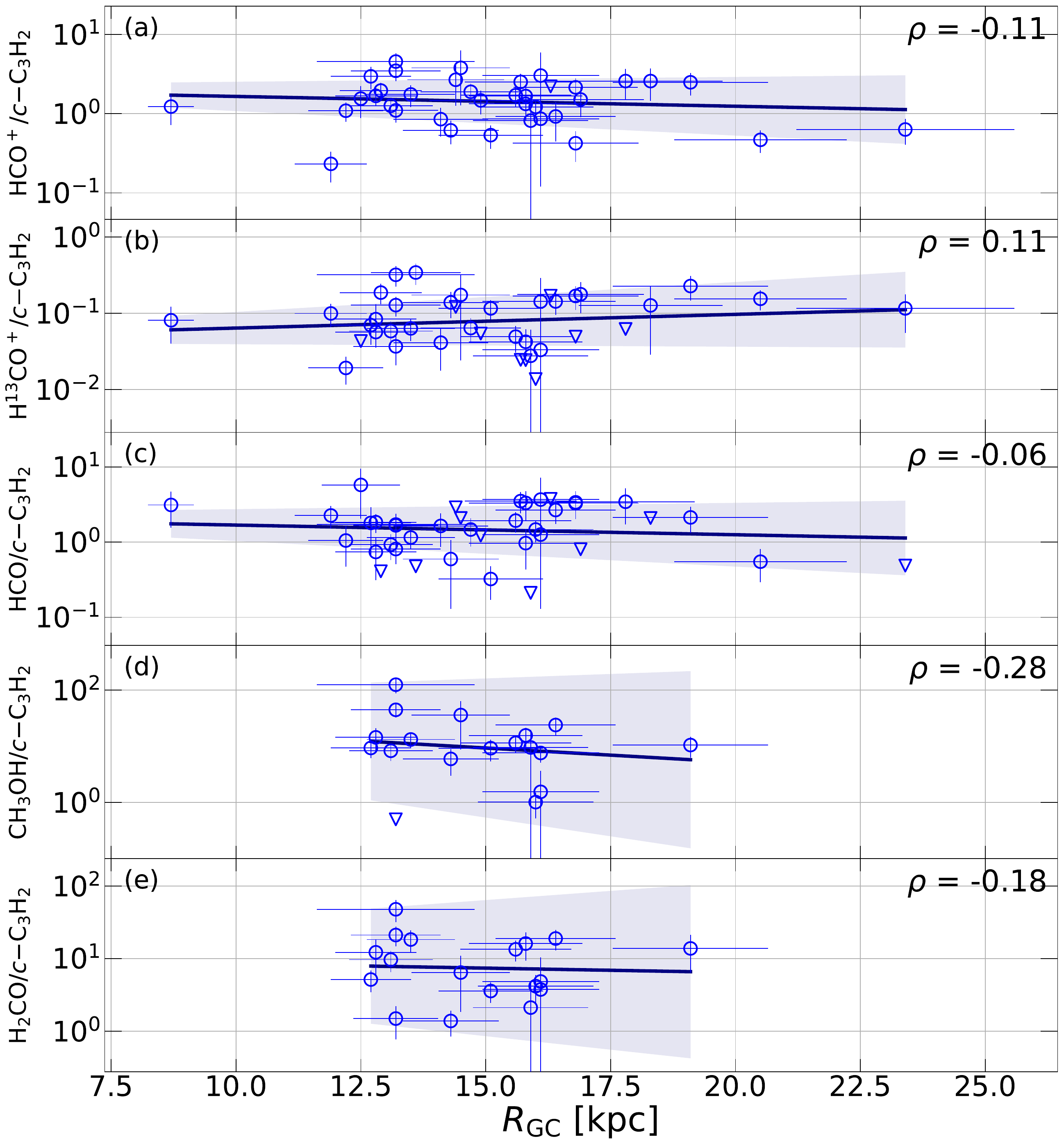}
            \caption{Galactocentric gradients of molecular ratios. The plot shows the trend of the ratios between CO molecule products (i.e., HCO$^+$, H$^{13}$CO$^+$, HCO, CH$_3$OH, and H$_2$CO) and \textit{c}-C$_3$H$_2$, as a function of galactocentric radius ($R_\text{GC}$). The $1\sigma$ error bars over the slope of the gradients are displayed for each molecular fit. The upper limit values are represented with triangles. The linear fit results are shown as the dark blue lines. The $1\sigma$ error bars over the slope of the gradients are displayed for each fit. In the upper right side of each subplot, the Pearson correlation coefficient, $\rho$, is shown.}
            \label{fig:mol_ratios}
        \end{figure*}

        Considering the lack of direct CO measurements for our sample, we can use species containing C and O that are formed from CO, such as HCO$^+$, H$^{13}$CO$^+$, HCO, H$_2$CO, and  CH$_3$OH. In Fig.~\ref{fig:mol_ratios}, the galactocentric gradients of these ratios are presented. No clear trends arise from the data, which are also characterized by $\left| \rho \right| < 0.2$ (except for methanol, for which we have $\rho\sim - 0.3$). 
        The ratio between H$^{13}$CO$^+$ and \textit{c}-C$_3$H$_2$ is the only one presenting a marginally increasing trend (in agreement with the formation pathways and the C/O abundance); however, it is too shallow to be considered reliable.
    \subsection{Caveats and uncertainties}
        \label{sect:caveats}
        The comparison between observations and chemical models by \cite{fontani2024chemout} for WB89-670 suggests that the methanol could be sub-thermally excited: the kinetic temperatures predicted by chemical modeling are in the $\sim20-40~$~K range, while the excitation temperatures are on the order of $\sim 10$~K. 
        Therefore, an important caveat arises from the assumed $T_\text{ex}$ for the species for which we cannot estimate it directly from the observations.
        
        Furthermore, the sources are located in a range of heliocentric distances, $d$, between 3 and 15.1~kpc, which implies that the linear scale probed is not the same in all targets. 
        \cite{fontani2024chemout} demonstrated the presence of a high differentiation at smaller angular scales in WB89-670, and hence the average column densities derived in this work refer to linear scales with (eventually) different average conditions.
        However, this caveat is typical of source surveys, and in any case, we compare emissions arising from core envelopes in all targets.
    
        Some results may also be influenced by low statistics, particularly within certain $R_\text{GC}$ ranges, where the data characterizing a few sources have a bigger impact on the gradients. 
        This concern is more significant for formaldehyde and methanol, given their even lower sampling rates (only 18 molecular clumps of 39 are observed in H$_2$CO and CH$_3$OH spectral windows), over a smaller $R_\text{GC}$ range \citep{fontani2022chemoutII}.
    
        Column densities of species such as HCN or HCO$^+$ might be affected by large optical-depth effects in the observed lines, which could impact the derived Galactocentric gradients. 
        However, for HCN, the column densities were corrected by the line opacity (Table~\ref{tab:abund1}) obtained from the hyperfine structure of the observed transition, and both HCN and HCO$^+$ were observed also in their less abundant $^{13}$C-bearing isotopologues (see Appendix~\ref{app:isotop} for the gradient of H$^{13}$CN).
        Therefore, a gradient can also be inferred for these species using these isotopologues, expected to be optically thin.
    
        Moreover, in the comparison between molecular and elemental trends, a further caveat comes from the fact that the elemental abundance gradients are extrapolated in the extreme OG ($R_\text{GC} \gtrsim 16$~kpc), where observations are sparse and thus the gradients are poorly constrained, and could be different from those at $R_\text{GC} \leq 16$~kpc. 
        Indeed, recent studies have reported evidence of a flattening of the metallicity gradient at large galactocentric radii, which could influence the observed molecular abundances and potentially explain the unexpectedly high chemical complexity in the outskirts of the Galaxy \citep{donor2020open, myers2022open, magrini2023gaia, da2023oxygen}.

\section{Conclusions and outlooks}
    \label{sect:6}
    In this work, we studied the chemical composition of 35 star-forming regions in the OG ($9\leq R_\text{GC}\leq24$~kpc), using IRAM 30 m telescope data from the CHEMOUT project. We identified 39 molecular clumps and analyzed the molecular species detected in the greatest number of targets across our sample; i.e., HCN (39/39),  HCO$^+$ (39/39), \textit{c}-C$_3$H$_2$ (38/39), H$^{13}$CO$^+$ (30/39), HCO (29/39), and SO (25/39). We estimated their column densities and calculated the fractional abundances of these species with respect to H$_2$. We also updated the abundances of H$_2$CO (18/18) and CH$_3$OH (17/18) from \cite{fontani2022chemoutII}, using the H$_2$ column densities obtained from the new NIKA2 continuum maps.
    
    The fractional abundances of the studied species with respect to H$_2$ estimated assuming a gas-to-dust ratio dependent on $R_\text{GC}$ \citep{giannetti2017galactocentric} show a decreasing trend with the galactocentric radius. For HCN, HCO$^+$, \textit{c}-C$_3$H$_2$, HCO, SO, H$_2$CO, and CH$_3$OH the abundances scale as the elemental abundance of carbon ([C/H]) between the solar circle and $\sim24~$kpc. This indicates that the efficiency in the formation of these molecules, which scales with the availability of the parent element (i.e., C), is at least as high as that found in the local Galaxy. 
    Instead, the decreasing rate of SO is higher than that of [S/H], indicating a lower efficiency in the formation of this species in the OG.
    Conversely, the decreasing rate of H$^{13}$CO$^+$ presents the puzzling result that it is lower than that of [$^{13}$C/H], resulting in a higher formation efficiency in the OG compared to the inner Galaxy.
    
    The galactocentric gradients of the abundances indicate how the metallicity of the environment, which scales as effectively with $R_\text{GC}$, does not affect the formation efficiency of most of the molecules studied, even for so-called prebiotic molecules such as formaldehyde.
    Metallicity appears to act mostly as a scaling factor for the molecular abundances relative to the elemental ones. 
    The role of metallicity in the efficiency of molecule formation may become evident when studying molecules more complex than methanol, and/or it could be hidden by other environmental factors such as the radiation field, gas temperature, or cosmic ionization rate, which may counterbalance its effect.
    These results confirm (after the previous papers of the CHEMOUT series) that the presence of organic molecules and tracers of protostellar activity is ubiquitous in the low-metallicity environment of the OG.\\
    
    The next step for the CHEMOUT project is to infer a more precise kinetic temperature of the sources to better estimate the column densities from the line analysis. 
    Determining this parameter will also help to reduce the space parameters on a chemical modeling analysis \citep[e.g., using UCLCHEM;][]{holdship2017uclchem}, which is crucial to inferring the environmental differences of the star-forming regions in the different areas of the Galaxy and understanding the parameters driving the chemistry \citep{vermarien2025understanding}. The comparison between high-resolution observations and chemical models is crucial to disentangling the several contributions to these formation processes.
    
    It is furthermore crucial to enhance the statistics acquired as much as we can. In fact, given the scarcity of star-forming regions in the extreme OG, the data we have for $R_\text{GC}\gtrsim20$~kpc weigh significantly on the overall trends. Increasing the observed sample is essential to constraining the gradients obtained and securing more precise results over the ongoing chemistry in this region of the Milky Way.

\begin{acknowledgements}
    We thank the anonymous referee for providing valuable comments and suggestions that improved the scientific content of this paper.
    F.F. is grateful to the IRAM 30 m staff for their precious help during the observations. L.C. and V.M.R acknowledge support from grant no. PID2022-136814NB-I00 by the Spanish Ministry of Science, Innovation and Universities/State Agency of Research MICIU/AEI/10.13039/501100011033 and by ERDF, UE. V.M.R. also acknowledges support from the grant RYC2020-029387-I funded by MICIU/AEI/10.13039/501100011033 and by "ESF, Investing in your future", and from the Consejo Superior de Investigaciones Cient{\'i}ficas (CSIC) and the Centro de Astrobiolog{\'i}a (CAB) through the project 20225AT015 (Proyectos intramurales especiales del CSIC); and from the grant CNS2023-144464 funded by MICIU/AEI/10.13039/501100011033 and by “European Union NextGenerationEU/PRTR”. The research leading to these results has received funding from the European Union's Horizon 2020 research and innovation programme under grant agreement nos. 730562 and 101004719 (ORP: \url{https://www.orp-h2020.eu}). SV and GV acknowledge support from the European Research Council (ERC) Advanced grant MOPPEX 833460. A.S.-M.\ acknowledges support from the RyC2021-032892-I grant funded by MCIN/AEI/10.13039/501100011033 and by the European Union `Next GenerationEU’/PRTR, as well as the program Unidad de Excelencia Mar\'ia de Maeztu CEX2020-001058-M, and support from the PID2023-146675NB-I00 (MCI-AEI-FEDER, UE).
\end{acknowledgements}

\bibliographystyle{aa} 
\bibliography{bibliography} 
    \begin{appendix} 
        \section{The CHEMOUT sample}
            Table~\ref{tab:sources} presents the observed sample together with its main properties.
            \begin{table*}
                \centering
                \caption{List of molecular clumps analyzed in this paper.}
                \begin{tabular}{lcccccccc}
                    \hline
                    \hline
                    \noalign{\smallskip}
                    \multirow{2}{*}{Source \tablefootmark{a}} & R.A. & Dec & $\ell$ & $V_\text{LSR}$\tablefootmark{b} & $R_\text{GC}$\tablefootmark{c} & $d$\tablefootmark{c} & $N_{\text{H}_2}$\tablefootmark{d} & $T_\text{ex}$\tablefootmark{e}\\
                    & (J2000) & (J2000) & (deg) & (km s$^{-1}$) & (kpc) & (kpc) & ($\times 10^{23}~\text{cm}^{-2}$) & (K) \\
                    \noalign{\smallskip}
                    \hline 
                    \noalign{\smallskip}
                    19383+2711 & \multirow{2}{*}{19:40:22.1} & \multirow{2}{*}{27:18:33} & \multirow{2}{*}{62.58} & -70.2 & 13.2 (0.9) & 14.8 (1.0) & \multirow{2}{*}{2.25 (0.19)} & ... \\
                    19383+2711-b &  &  &  & -65.6 & 12.7 (0.8) & 14.2 (0.9) & & 12.4 (0.8) \\
                    19423+2541 & 19:44:23.2 & 25:48:40 & 61.72 & -72.58 & 13.5 (0.9) & 15.3 (1.0) & 2.9 (0.4) & 9.9 (0.4)\\
                    19489+3030 & 19:50:53.2 & 30:38:09 & 66.61 & -69.29 & 12.9 (0.8) & 13.7 (0.9) & 0.57 (0.06) & ... \\
                    19571+3113 & \multirow{2}{*}{19:59:08.5} & \multirow{2}{*}{31:21:47} & \multirow{2}{*}{68.15} & -66.2 & 12.5 (0.8) & 13.0 (0.8) & \multirow{2}{*}{0.79 (0.08)} & ... \\
                    19571+3113-b &  &  &  & -61.7 & 12.2 (0.8) & 12.5 (0.8) &  & ... \\
                    20243+3853 & 20:26:10.8 & 39:03:30 & 77.61 & -73.21 & 12.8 (0.8) & 11.7 (0.8) & 0.77 (0.07) & ... \\
                    WB89-002  & 20:37:22.3 & 47:13:54 & 85.41 & -2.83 & 8.7 (0.5) & 3.1 (1.6) & ... & ... \\
                    WB89-006 & \multirow{2}{*}{20:42:58.2}  & \multirow{2}{*}{47:35:35} & \multirow{2}{*}{86.27} & -92.3 & 14.5 (1.0) & 12.4 (0.9) & \multirow{2}{*}{0.79 (0.08)} & 9.8 (0.7)\\
                    WB89-006-b  & &  &  & -90.3 & 14.3 (1.0) & 12.2 (0.9) &  & 9.4 (1.0)\\
                    WB89-014  & 20:52:07.8 & 49:51:28 & 88.99 & -96.0 & 14.9 (1.0) & 12.5 (1.0) & 0.25 (0.03) & ... \\
                    WB89-031  & 21:04:18.0 & 46:53:10 & 88.06 & -88.89 & 14.1 (1.0) & 11.7 (0.9) & 0.38 (0.03) & ... \\
                    WB89-035  & 21:05:19.7 & 49:15:59 & 89.94 & -77.56 & 13.1 (0.8) & 10.1 (0.8) & 0.39 (0.05) & 11.7 (1.3) \\
                    WB89-040  & 21:06:50.0 & 50:02:09 & 90.68 & -62.38 & 11.9 (0.7) & 8.3 (0.7) & 0.13 (0.02) & ... \\
                    WB89-060  & 21:15:56.0 & 54:43:33 & 95.05 & -83.7 & 13.6 (0.9) & 10.1 (0.8) & 1.7 (0.2) & ... \\
                    WB89-076  & 21:24:29.0 & 53:45:35 & 95.25 & -97.07 & 15.1 (1.1) & 11.8 (0.9) & 1.03 (0.10) & 7.5 (1.1) \\
                    WB89-080  & 21:26:29.1 & 53:44:11 & 95.44 & -74.1 & 12.8 (0.8) & 8.9 (0.8) & 0.51 (0.04) & 9.2 (0.9) \\
                    WB89-083  & 21:27:47.7 & 54:26:58 & 96.08 & -93.76 & 14.7 (1.0) & 11.2 (0.9) & 0.191 (0.015) & ... \\
                    WB89-152  & 22:05:15.4 & 60:48:41 & 104.0 & -88.5 & 14.4 (1.0) & 9.8 (0.9) & ... & ... \\
                    WB89-283  & 23:32:23.8 & 63:33:18 & 114.3 & -94.69 & 15.8 (1.1) & 10.4 (0.9) & 0.36 (0.04) & 14.9 (1.1) \\
                    WB89-288  & 23:36:08.1 & 62:23:48 & 114.3 & -101.0 & 16.8 (1.2) & 11.5 (1.0) & 0.54 (0.04) & ... \\
                    WB89-315 & 00:05:53.8 & 64:05:17 & 118.0 & -95.1 & 16.3 (1.2) & 10.7 (1.0) & ... & ... \\
                    WB89-379  & 01:06:59.9 & 65:20:51 & 124.6 & -89.16 & 16.4 (1.2) & 10.2 (1.0) & 1.37 (0.16) & 9.3 (0.9) \\
                    WB89-380 & \multirow{2}{*}{01:07:50.9} & \multirow{2}{*}{65:21:22} & \multirow{2}{*}{124.6} & -87.4 & 16.1 (1.2) & 9.8 (0.9) & \multirow{2}{*}{5.0 (0.6)} & 13 (4) \\
                    WB89-380-b  &  &  &  & -86.0 & 15.9 (1.2) & 9.6 (0.9) &  & 11.4 (0.5) \\
                    WB89-391  & 01:19:27.1 & 65:45:44 & 125.8 & -86.1 & 16.1 (1.2) & 9.7 (0.9) & 1.19 (0.13) & 9.7 (0.8) \\
                    WB89-399  & 01:45:39.4 & 64:16:00 & 128.8 & -82.15 & 16.0 (1.2) & 9.4 (0.9) & 1.75 (0.17) & 11 (3) \\
                    WB89-437  & 02:43:29.0 & 62:57:08 & 135.3 & -72.14 & 13.2 (0.9) & 5.9 (0.7) & 2.1 (0.2) & 11.8 (0.4) \\
                    WB89-440  & 02:46:07.3 & 62:46:31 & 135.6 & -71.88 & 15.7 (1.1) & 8.6 (0.9) & 0.69 (0.06) & ... \\
                    WB89-501  & 03:52:27.6 & 57:48:34 & 145.2 & -58.43 & 15.6 (1.1) & 8.0 (0.8) & 1.20 (0.13) & 11.0 (1.2) \\
                    WB89-529  & 04:06:25.5 & 53:21:49 & 149.6 & -59.8 & 17.8 (1.4) & 10.1 (1.0) & 1.17 (0.09) & ... \\
                    WB89-572  & 04:35:58.3 & 47:42:58 & 156.9 & -47.4 & 18.3 (1.4) & 10.3 (1.1) & 1.17 (0.13) & ... \\
                    WB89-621  & 05:17:13.4 & 39:22:15 & 168.1 & -25.68 & 13.2 (1.6) & 5.0 (1.2) & 1.8 (0.2) & 10.3 (0.4) \\
                    WB89-640  & 05:25:40.7 & 41:41:53 & 167.1 & -24.93 & 16.8 (1.3) & 8.6 (0.9) & 2.9 (0.2) & ... \\
                    WB89-670  & 05:37:41.9 & 36:07:22 & 173.0 & -17.65 & 23.4 (2.1) & 15.1 (1.6) & 1.34 (0.17) & ... \\
                    WB89-705  & 05:47:47.6 & 35:22:01 & 174.7 & -12.2 & 20.5 (1.7) & 12.0 (1.3) & 0.68 (0.08) & ... \\
                    WB89-789  & 06:17:24.3 & 14:54:37 & 195.8 & 34.25 & 19.1 (1.6) & 11.0 (1.1) & 4.1 (0.4) & 9.7 (1.0) \\
                    WB89-793  & 06:18:41.7 & 15:04:52 & 195.8 & 30.5 & 16.9 (1.3) & 8.7 (0.9) & 0.99 (0.11) & ... \\
                    WB89-898  & 06:50:37.3 & –05:21:01 & 217.6 & 63.5 & 15.8 (1.1) & 8.4 (0.9) & ... & ...\\
                    \noalign{\smallskip}
                    \hline
                \end{tabular}
                \label{tab:sources}
                \tablefoot{
                    \tablefoottext{a}{The sources labeled with a "-b" are related to secondary velocity features detected in the spectra, likely due to a second molecular clump within the telescope beam.} 
                    \tablefoottext{b}{Local standard of rest velocities estimated from the velocity peak of the \textit{c}-C$_3$H$_2$ $J_{K_a,K_c} = 2_{1,2} - 1_{0,1}$ line, from \cite{fontani2022chemout}. For the source WB89-315, being undetected in \textit{c}-C$_3$H$_2$, we use the $V_\text{LSR}$ given in \cite{blair2008formaldehyde}.}
                    \tablefoottext{c}{Galactocentric distances and heliocentric distances, respectively, based on the rotation curve of \cite{russeil2017milky}, and the $V_\text{LSR}$ values shown in this table. For WB89-437 and WB89-621, we assume the more accurate distances estimated through maser parallax measurements in \cite{hachisuka2015parallaxes}. Their kinematic galactocentric distances are 15.7 and 18.9~kpc, respectively.}
                    \tablefoottext{d}{Derived from the NIKA2 continuum maps as described in Sect.~\ref{sect:continuummethods}.} 
                    \tablefoottext{e}{Excitation temperature of CH$_3$OH, derived from the updated fits of the data presented in \cite{fontani2022chemoutII} (see Sect.~\ref{sect:updates}), adopted to fit the other species.}
                        }                             
            \end{table*}
        \section{Millimeter continuum maps and H$_2$ column densities}
            \label{NIKA2}
\begin{figure*}[h]
                    \centering
                    \includegraphics[width=0.90\linewidth]{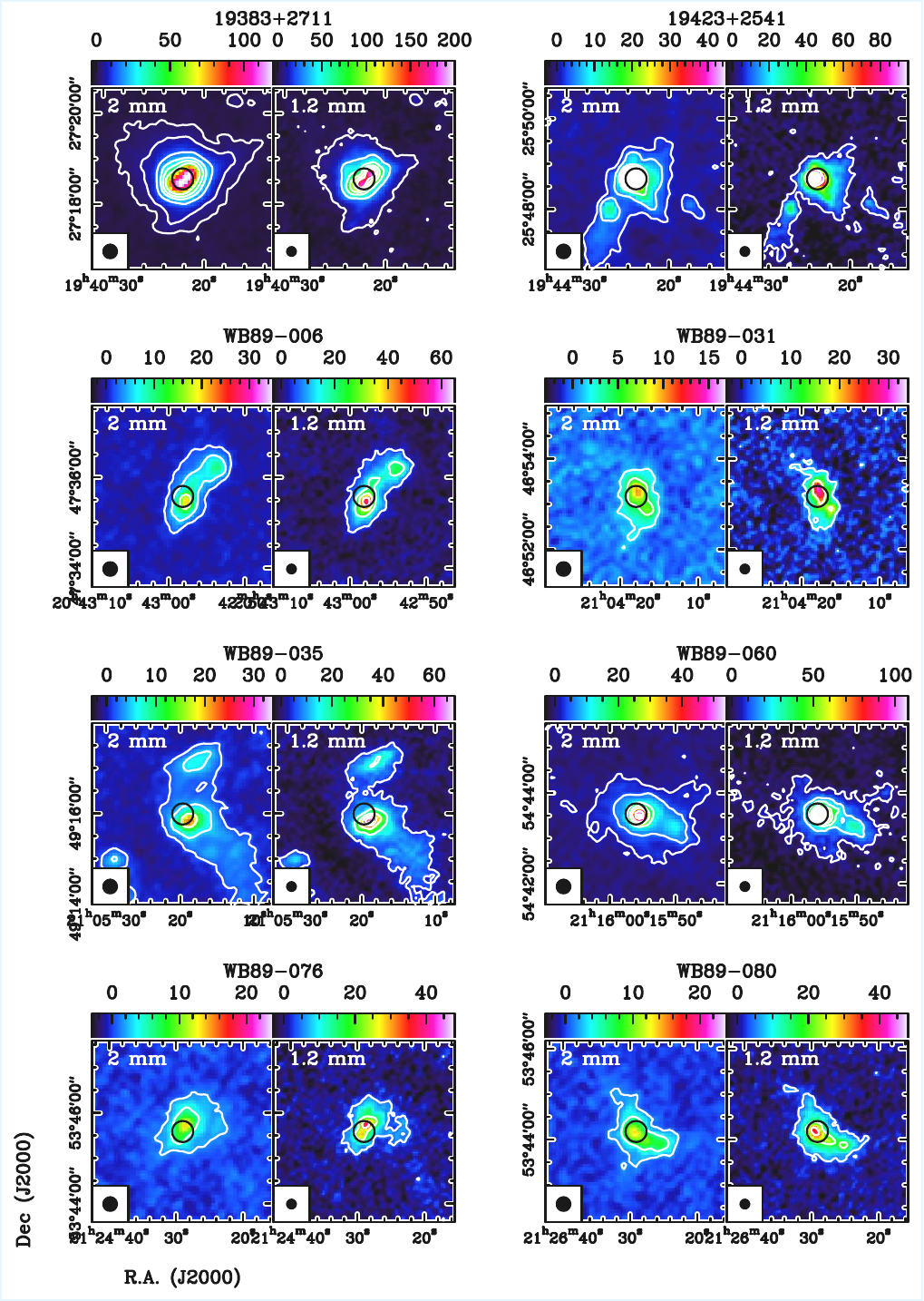}
                    \caption{\label{fig:NIKA2-1} Preliminary reduced NIKA2 images (color-scale in mJy/beam) at 2.0~mm (150~GHz) and 1.2~mm (260~GHz) of 8 out of the 31 CHEMOUT sources observed (Fontani et al. in prep.). The contours start from the 5$\sigma$ rms level in each map. The NIKA2 angular resolution is shown in the bottom-left corner of each frame. The black circles on the maps represent the 28$''$ beam, centered on the position targeted for the molecular line observations analyzed in this paper.}
            \end{figure*}

\begin{figure*}[h]
                    \centering
                    \includegraphics[width=0.93\linewidth]{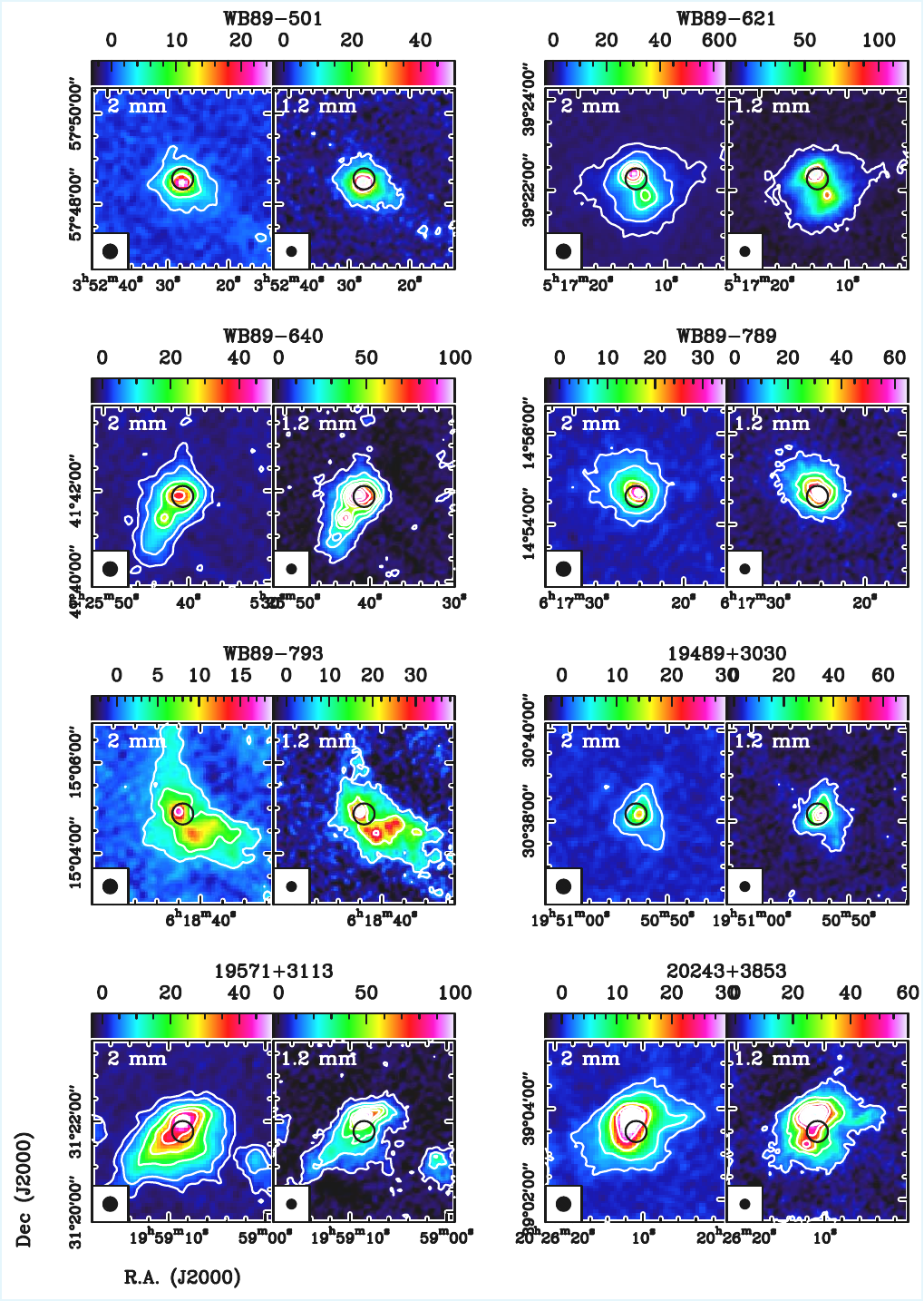}
                    \caption{\label{fig:NIKA2-2} Same as Fig.~\ref{fig:NIKA2-1} for 8 additional targets.}
            \end{figure*}

\begin{figure*}[h]
                    \centering
                    \includegraphics[width=0.93\linewidth]{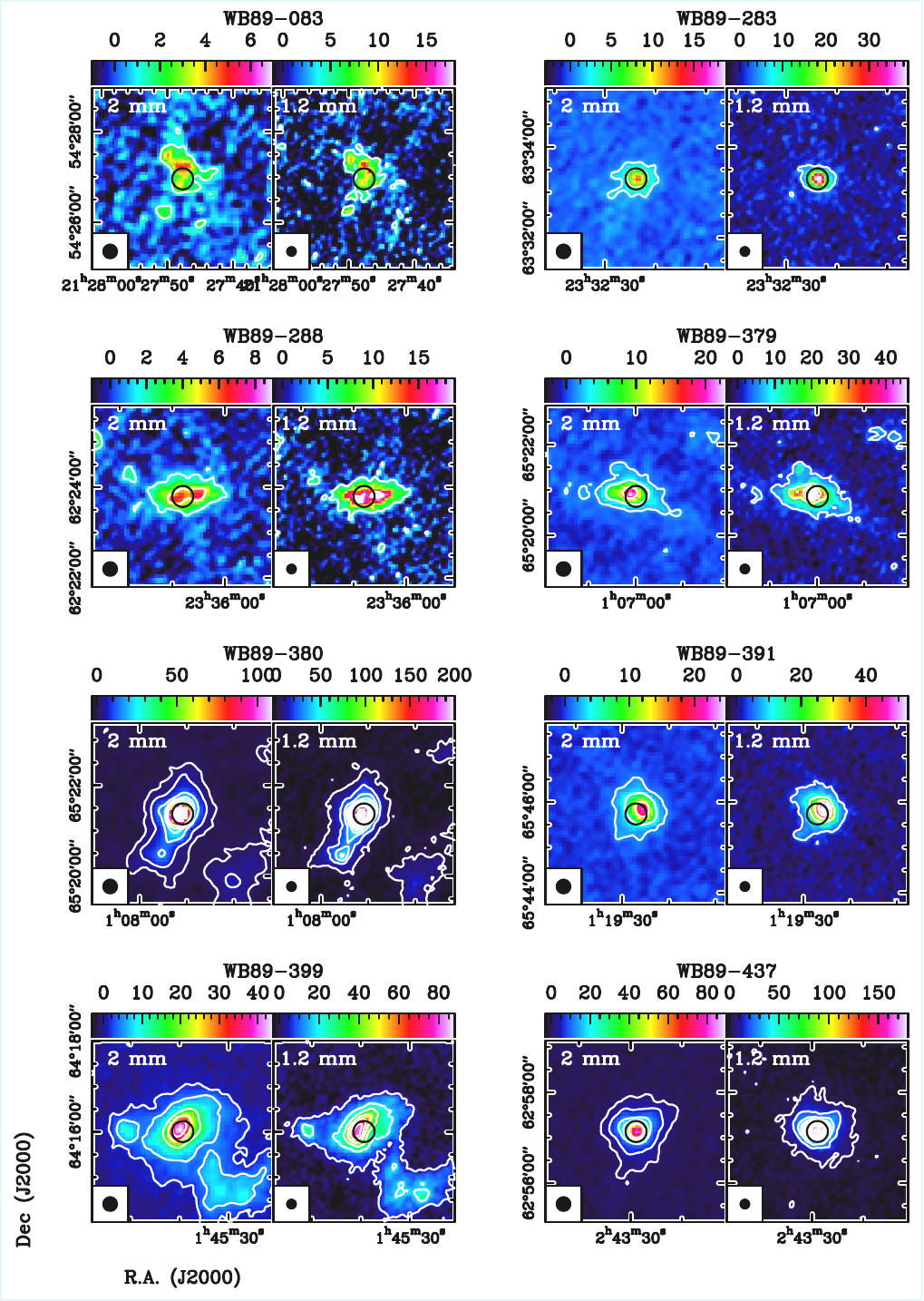}
                    \caption{\label{fig:NIKA2-3} Same as Fig.~\ref{fig:NIKA2-1} for 8 additional targets.}
            \end{figure*}

\begin{figure*}[h]
                    \centering
                    \includegraphics[width=0.93\linewidth]{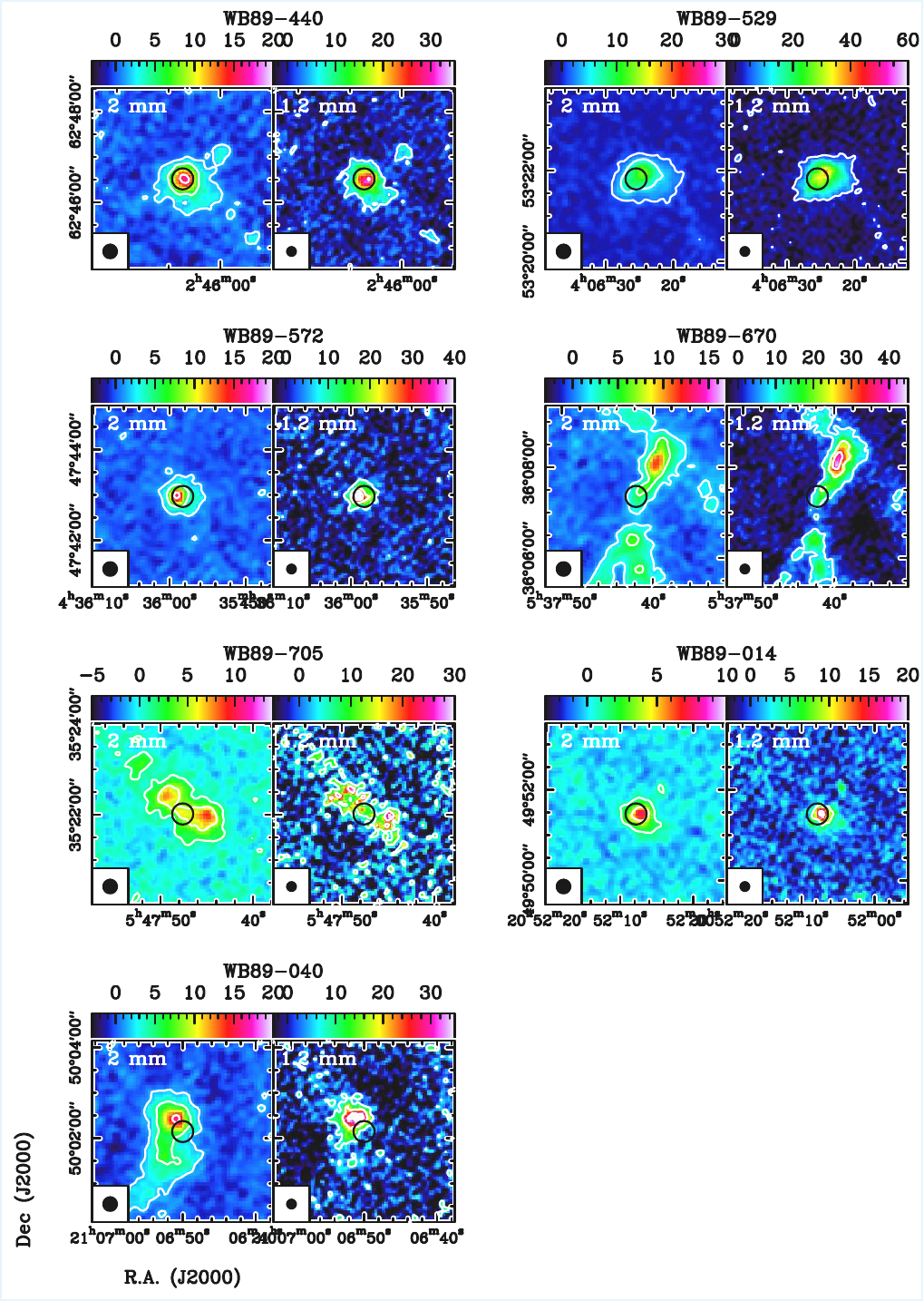}
                   \caption{\label{fig:NIKA2-4} Same as Fig.~\ref{fig:NIKA2-1} for the remaining 7 targets.}
            \end{figure*}

            In this appendix, we show the millimeter continuum maps obtained with the IRAM 30m bolometer NIKA2 (Figs.~\ref{fig:NIKA2-1}-\ref{fig:NIKA2-4}).
            We observed 31 out of the 35 CHEMOUT targets.            
            A detailed analysis of the maps will be presented in a forthcoming paper (Fontani et al. in prep.). 
            
\begin{figure*}
                \centering
                \includegraphics[width=\linewidth]{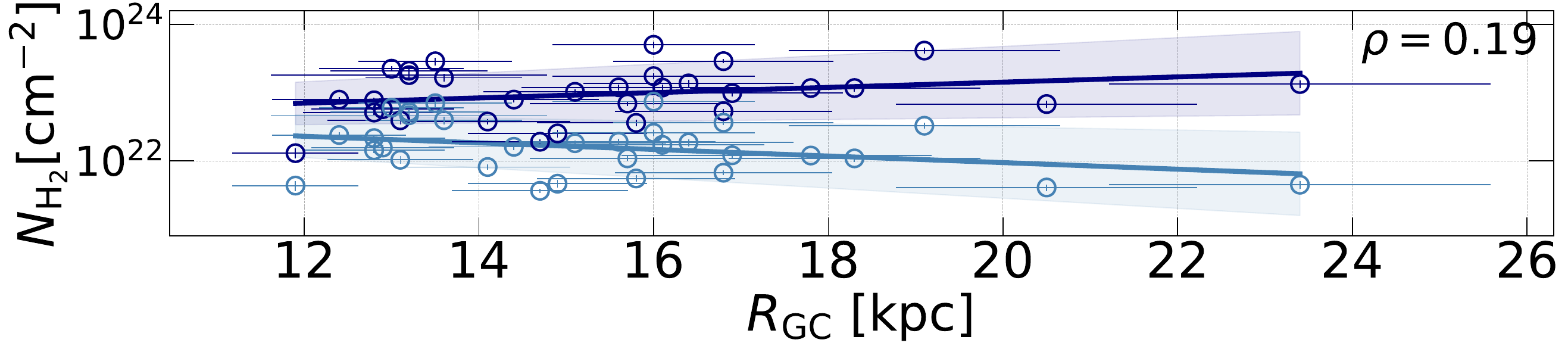}
                \caption{Gradients of column densities of H$_2$, $N_{\text{H}_2}$, as a function of galactocentric radius ($R_\text{GC}$). The light blue dataset represents the column densities estimated using a constant gas-to-dust ratio ($\gamma=100$), while the dark blue dataset shows the column densities calculated with a gas-to-dust ratio function of $R_\text{GC}$ from \cite{giannetti2017galactocentric}. Both datasets have been fitted with a linear regression. The $1\sigma$ error bars over the slope of the gradients are displayed for each fit. In the upper-right side of the plot, the Pearson correlation coefficient, $\rho$, is shown (only for the dark blue dataset).}
                \label{fig:densities2}
            \end{figure*}

            In Table~\ref{tab:fluxes}, all the parameters derived from the continuum maps, and used to estimate the H$_2$ column densities, are presented. The fluxes have been extracted from the NIKA2 maps using a 28~$''$ beam, matching the widest beam size of the spectral data centered at the same position from which we observed the IRAM 30m spectra. The spectral index $\beta$ is estimated using the following equation:
            \begin{equation}
                \beta = \frac{\log(\frac{F_{\nu_2}}{F_{\nu_1}}\frac{B_{\nu_1}(T_\text{d})}{B_{\nu_2}(T_\text{d})})}{\log\frac{\nu_2}{\nu_1}},
            \end{equation}
            where $\nu_2$ and $\nu_1$, are 150 and 260~GHz (i.e., the frequencies corresponding to 2.0 and 1.2~mm), and $T_\text{d}$ is the assumed dust temperature (see Sect.~\ref{sect:continuummethods}).
            In Fig.~\ref{fig:densities2}, the galactocentric trends of H$_2$ are shown. The dataset estimated with a non-constant gas-to-dust ratio $\gamma=\gamma(R_\text{GC})$ \citep{giannetti2017galactocentric} exhibits a positive trend with increasing values of $R_\text{GC}$, and a weak, positive correlation ($\rho = 0.19$), which is due to the increasing $\gamma$ (Sect.~\ref{sect:continuummethods}).
            
            \begin{table}[H]
                \centering
                \setlength{\tabcolsep}{3pt}
                \renewcommand{\arraystretch}{0.92} 
                \caption{Parameters (i.e., fluxes, gas-to-dust ratios $\gamma$, and spectral index $\beta$) used in the calculation of the H$_2$ column densities.} 
                \begin{tabular}{lcccc}
                    \hline
                    \hline
                    \noalign{\smallskip}
                    Source & $\gamma$ & $\beta$ & $F_{2.0\text{mm}}$ & $F_{1.2\text{mm}}$\\
                     & & & (mJy) & (mJy) \\
                    \noalign{\smallskip}
                    \hline
                    \noalign{\smallskip}
                    19383+2711 & 372 & 0.417 & 178 (5) & 519 (17) \\
                    19423+2541 & 412 & 1.051 & 127 (10) & 499 (38) \\
                    19489+3030 & 365 & 2.248 & 19 (1) & 148 (7) \\
                    19571+3113 & 330 & 0.182 & 70 (2) & 175 (9) \\
                    20243+3853 & 358 & 0.875 & 47 (2) & 172 (7) \\
                    WB89-006 & 493 & 1.826 & 20 (1) & 119 (6) \\
                    WB89-014 & 545 & 1.329 & 8.3 (0.4) & 39 (3) \\
                    WB89-031 & 464 & 1.744 & 12.2 (0.5) & 72 (3) \\
                    WB89-035 & 380 & 1.761 & 18 (1.6) & 111 (9) \\
                    WB89-040 & 299 & 0.514 & 11 (1) & 33 (4) \\
                    WB89-060 & 420 & 1.836 & 57 (4) & 354 (28) \\
                    WB89-076 & 567 & 1.978 & 14.6 (0.7) & 85 (4) \\
                    WB89-080 & 358 & 1.819 & 16.8 (0.6) & 98 (3) \\
                    WB89-083 & 523 & 1.947 & 5.0 (0.2) & 33 (1.0) \\
                    WB89-283 & 653 & 1.711 & 13.1 (0.7) & 83 (5) \\
                    WB89-288 & 797 & 1.955 & 9.2 (0.2) & 61 (1.6) \\
                    WB89-379 & 736 & 2.063 & 20 (1.5) & 134 (9) \\
                    WB89-380 & 679 & 0.289 & 242 (18) & 664 (42) \\
                    WB89-391 & 693 & 2.021 & 20 (1.6) & 133 (8) \\
                    WB89-399 & 679 & 0.922 & 57 (2) & 216 (10) \\
                    WB89-437 & 403 & 1.802 & 87 (5) & 546 (34) \\
                    WB89-440 & 640 & 1.037 & 22 (1) & 88 (3) \\
                    WB89-501 & 627 & 1.893 & 28 (1.6) & 181 (10) \\
                    WB89-529 & 974 & 2.044 & 15.8 (0.5) & 110 (3) \\
                    WB89-572 & 1077 & 1.928 & 15 (1) & 98 (6) \\
                    WB89-621 & 388 & 1.872 & 63 (4) & 394 (28) \\
                    WB89-640 & 797 & 1.808 & 53 (1.2) & 324 (7) \\
                    WB89-670 & 2991 & 2.010 & 6.0 (0.4) & 41 (3) \\
                    WB89-705 & 1673 & 1.026 & 8.3 (0.3) & 33 (2) \\
                    WB89-789 & 1264 & 1.842 & 41 (2) & 247 (13) \\
                    WB89-793 & 813 & 1.641 & 19 (1.0) & 106 (6) \\
                    \noalign{\smallskip}
                    \hline
                \end{tabular}
                \label{tab:fluxes}
            \end{table}

        \section{Line-fitting results}
            \label{app:parameters}
            In this appendix, we present the results of the line fit from \textsc{MADCUBA}.
            Tables \ref{tab:abund1} and \ref{tab:abund2} present the best-fit parameters from \textsc{MADCUBA} for HCN, HCO$^+$, \textit{c}-C$_3$H$_2$, H$^{13}$CO$^+$, HCO, and SO. The fit results for H$_2$CO and CH$_3$OH have already been presented in \cite{fontani2022chemoutII}. The upper-limits values are also estimated for the undetected transitions. For the (self-) absorbed lines, no best-fit parameters are given.

\begin{figure*}
                \centering
                \includegraphics[width=\linewidth]{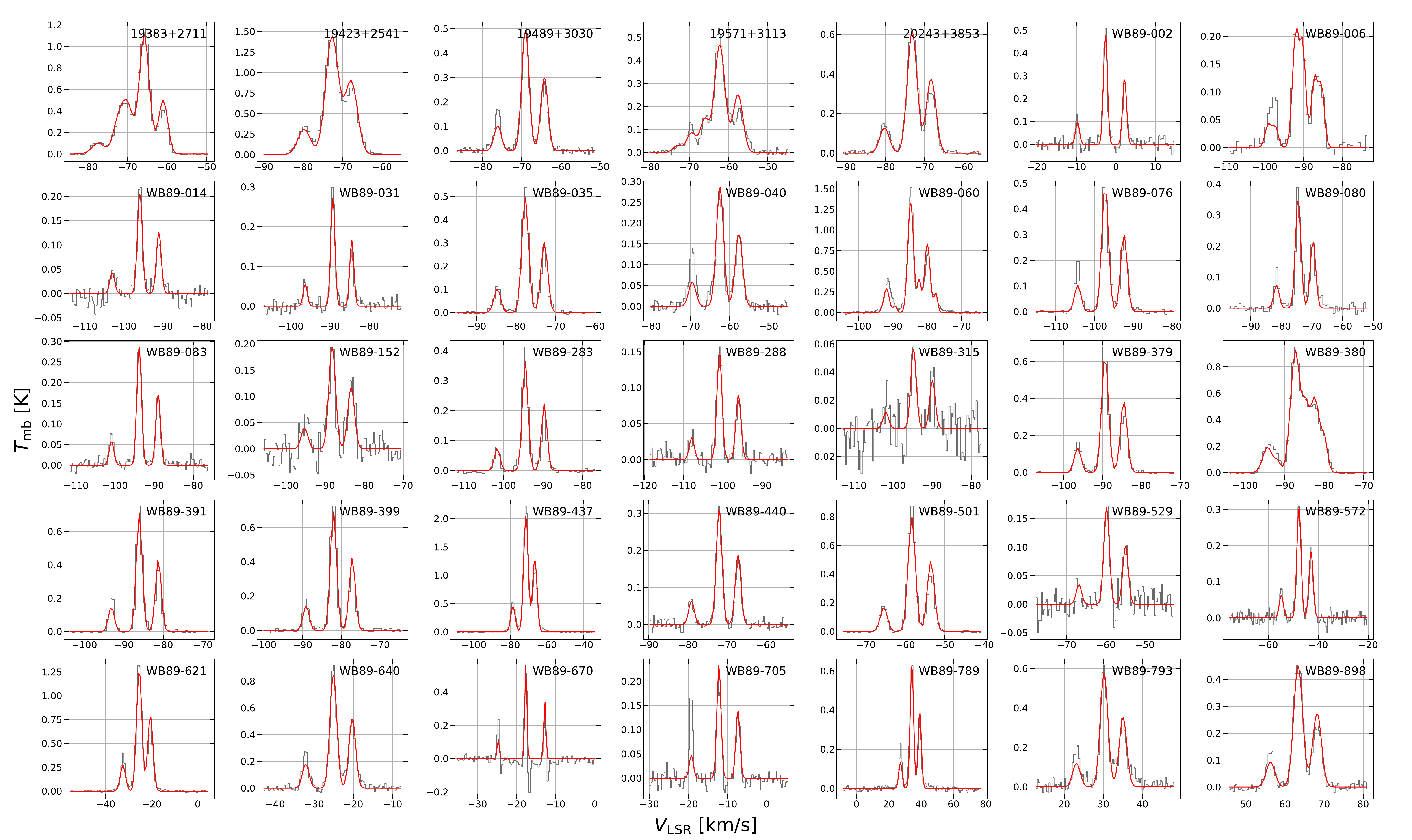}
                \caption{Spectra of HCN $J=1-0$ ($F=1-1; F=2-1; F=0-1$). The observed spectra (grey histograms) are superimposed on the synthetic Gaussian fits obtained with \textsc{MADCUBA}. The red lines represent the fit of the detected transitions, and the blue lines represent the upper limits of the non-detected transitions.}
                \label{fig:spectra1}
            \end{figure*}

\begin{figure*}
                \centering
                \includegraphics[width=\linewidth]{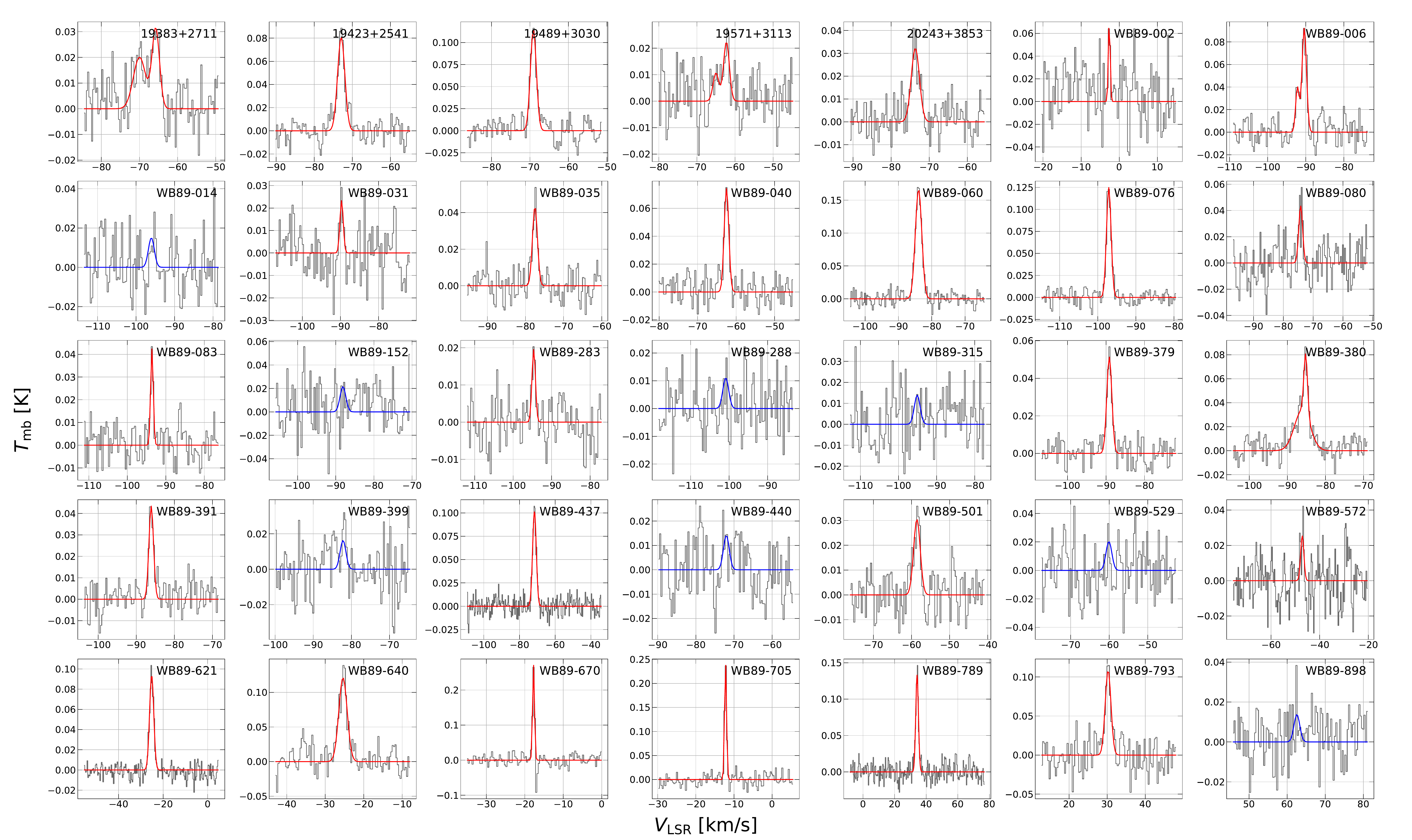}
                \caption{Same as Fig.~\ref{fig:spectra1} for H$^{13}$CO$^+$ $J=1-0$.}
                \label{fig:spectra2}
            \end{figure*}

\begin{figure*}
                \centering
                \includegraphics[width=\linewidth]{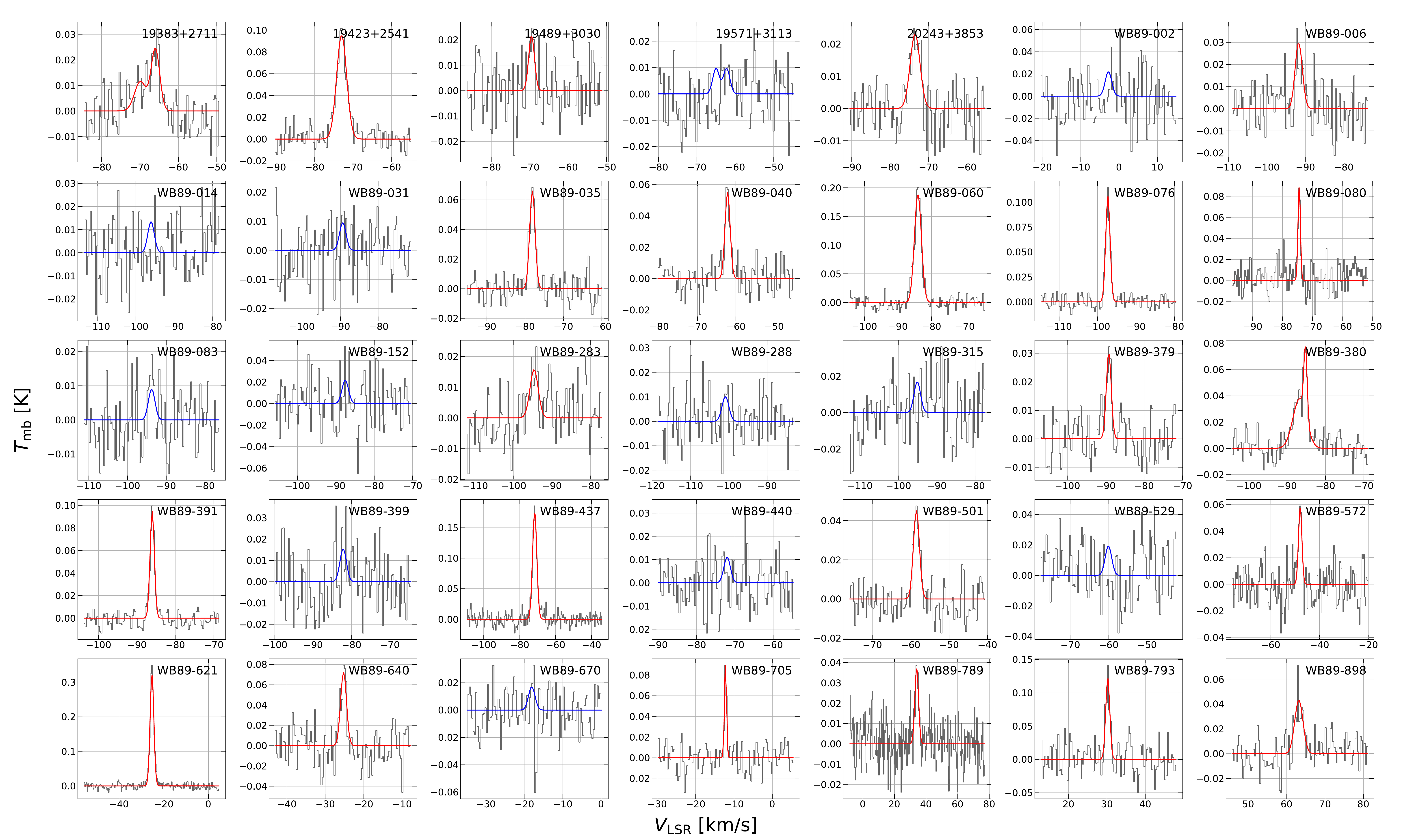}
                \caption{Same as Fig.~\ref{fig:spectra1} for SO $N=2-1, J=2-1$.}
                \label{fig:spectra3}
            \end{figure*}

            In Figs.~\ref{fig:spectra1}, \ref{fig:spectra2}, and \ref{fig:spectra3}, the observed spectra of the whole sample are shown with the Gaussian fits performed with \textsc{MADCUBA} for HCN, H$^{13}$CO$^+$, and SO. The plots showing the spectra of the remaining species are shown in \cite{fontani2022chemout} (\textit{c}-C$_3$H$_2$, HCO$^+$) and in \cite{fontani2022chemoutII} (HCO, H$_2$CO, CH$_3$OH).
          
            \begin{sidewaystable*}
                \centering
                \tiny
                \setlength{\tabcolsep}{1.3pt}
                \caption{HCN, HCO$^+$, \textit{c}-C$_3$H$_2$, and H$^{13}$CO$^+$ line parameters obtained from the fitting procedure. The values of each parameter are given with the error in parentheses. If the error is not shown, the parameter was fixed during the fit. The column density values have been scaled to a 28$''$ beam (Sect.~\ref{methods:fit}).}
                \begin{tabular}{lcccccccccccccccccccc}
                    \hline
                    \hline
                    \noalign{\smallskip}
                    & \multicolumn{5}{c}{HCN} & \multicolumn{1}{c}{} & \multicolumn{4}{c}{HCO$^+$} & \multicolumn{1}{c}{} & \multicolumn{4}{c}{\textit{c}-C$_3$H$_2$} & \multicolumn{1}{c}{} & \multicolumn{4}{c}{H$^{13}$CO$^+$}\\
                    \noalign{\smallskip}
                    \cline{2-6} \cline{8-11} \cline{13-16} \cline{18-21}
                    \noalign{\smallskip}
                    Source & $N_\text{tot}$ & $T_\text{ex}$ & $V_\text{LSR}$ & FWHM & $\tau_\nu$\tablefootmark{a} & & $N_\text{tot}$ & $T_\text{ex}$ & $V_\text{LSR}$ & FWHM & & $N_\text{tot}$ & $T_\text{ex}$ & $V_\text{LSR}$ & FWHM & & $N_\text{tot}$ & $T_\text{ex}$ & $V_\text{LSR}$ & FWHM\\
                    & ($\times 10^{13}$~cm$^{-2}$) & (K) & (km s$^{-1}$) & (km s$^{-1}$) & & & ($\times 10^{12}$~cm$^{-2}$) & (K) & (km s$^{-1}$) & (km s$^{-1}$) & & ($\times 10^{12}$~cm$^{-2}$) & (K) & (km s$^{-1}$) & (km s$^{-1}$) & & ($\times 10^{11}$~cm$^{-2}$) & (K) & (km s$^{-1}$) & (km s$^{-1}$) \\
                    \noalign{\smallskip}
                    \hline
                    \noalign{\smallskip}
                    19383+2711 & 0.70(0.11) & 10.7 & -70.36(0.10) & 3.8(0.2) & 0.067 (0.005) &  & 3.1(0.5) & 10.7 & -70.2(0.1) & 4.3(0.2) &  & 2.7(0.4) & 4.4(0.4) & -70.10(0.11) & 4.3(0.3) &  & 1.0(0.3) & 10.7 & -70.0(0.3) & 3.9(0.9) \\
                    19383+2711-b & 0.90(0.12) & 12.4 & -65.80(0.04) & 2.65(0.10) & 0.092 (0.004) &  & 4.4(0.6) & 12.4 & -65.80(0.04) & 2.80(0.08) &  & 1.4(0.2) & 6.2(0.5) & -65.5(0.1) & 3.0(0.2) &  & 1.0(0.3) & 12.4 & -65.7(0.2) & 2.3(0.4) \\
                    19423+2541 & 2.2(0.3) & 9.9 & -72.70(0.04) & 3.70(0.08) & 0.237 (0.001) &  & 6.6(1.0) & 9.9 & -72.800(0.015) & 3.00(0.06) &  & 3.5(0.5) & 6.1(0.4) & -72.60(0.06) & 3.60(0.14) &  & 2.2(0.4) & 9.9 & -72.90(0.08) & 2.2(0.2) \\
                    19489+3030 & 0.40(0.05) & 10.7 & -69.00(0.02) & 2.20(0.05) & 0.067 (0.003) &  & 3.0(0.3) & 10.7 & -69.100(0.009) & 2.60(0.03) &  & 1.4(0.2) & 10.7 & -69.30(0.04) & 1.96(0.10) &  & 2.7(0.4) & 10.7 & -69.10(0.05) & 1.80(0.12) \\
                    19571+3113 & 0.14(0.04) & 10.7 & -66.1(0.2) & 2.7(0.5) & 0.019 (0.005) &  & 0.9(0.2) & 10.7 & -65.90(0.05) & 1.90(0.13) &  & 0.60(0.15) & 10.7 & -66.10(0.11) & 1.7(0.3) &  & $\leq$ 0.2 & 10.7 & -66.2 & 1.8 \\
                    19571+3113-b & 0.50(0.08) & 10.7 & -62.40(0.08) & 2.9(0.2) & 0.056 (0.005) &  & 3.1(0.4) & 10.7 & -62.20(0.04) & 3.36(0.10) &  & 2.6(0.4) & 10.7 & -61.70(0.12) & 4.7(0.3) &  & 0.50(0.12) & 10.7 & -62.2(0.2) & 1.8 \\
                    20243+3853 & 0.70(0.08) & 10.7 & -73.10(0.03) & 2.90(0.05) & 0.084 (0.004) &  & 3.3(0.4) & 10.7 & -73.20(0.04) & 3.28(0.10) &  & 1.8(0.2) & 10.7 & -73.20(0.05) & 2.70(0.11) &  & 1.0(0.2) & 10.7 & -73.6(0.2) & 2.5(0.4) \\
                    WB89-002 & 0.20(0.03) & 10.7 & -2.70(0.02) & 1.20(0.05) & 0.066 (0.004) &  & 0.80(0.14) & 10.7 & -2.50(0.05) & 1.50(0.12) &  & 0.6(0.2) & 10.7 & -2.80(0.07) & 0.9(0.2) &  & 0.50(0.13) & 10.7 & -2.60(0.04) & 0.60(0.12) \\
                    WB89-006 & 0.20(0.06) & 9.8 & -92.0(0.2) & 2.0(0.4) & 0.032 (0.003) &  & 1.2(0.2) & 9.8 & -92.20(0.02) & 2.00(0.05) &  & 0.3(0.2) & 12.0(6.0) & -92.30(0.14) & 1.2(0.4) &  & 0.5(0.2) & 9.8 & -92.20(0.12) & 1.1(0.3) \\
                    WB89-006-b & 0.11(0.05) & 9.4 & -90.1(0.2) & 1.7(0.4) & 0.028 (0.003) &  & 0.74(0.10) & 9.4 & -89.60(0.02) & 1.50(0.05) &  & 1.1(0.2) & 7.3(0.8) & -90.30(0.04) & 1.50(0.11) &  & 1.5(0.3) & 9.4 & -90.40(0.06) & 1.40(0.14) \\
                    WB89-014 & 0.12(0.02) & 10.7 & -95.90(0.03) & 1.50(0.06) & 0.028 (0.002) &  & 1.00(0.12) & 10.7 & -95.90(0.02) & 1.80(0.04) &  & 0.60(0.13) & 10.7 & -96.00(0.09) & 1.6(0.2) &  & $\leq$ 0.3 & 10.7 & -96.0 & 1.8 \\
                    WB89-031 & 0.13(0.02) & 10.7 & -89.30(0.02) & 1.20(0.04) & 0.037 (0.002) &  & 0.70(0.13) & 10.7 & -89.400(0.012) & 1.10(0.04) &  & 0.70(0.15) & 10.7 & -88.90(0.08) & 1.6(0.2) &  & 0.30(0.11) & 10.7 & -89.70(0.14) & 1.0(0.3) \\
                    WB89-035 & 0.40(0.05) & 11.7 & -77.70(0.02) & 2.00(0.06) & 0.060 (0.002) &  & 2.0(0.2) & 11.7 & -77.600(0.008) & 1.80(0.02) &  & 1.5(0.2) & 4.7(0.5) & -77.60(0.04) & 1.50(0.08) &  & 0.9(0.2) & 11.7 & -77.50(0.07) & 1.5(0.2) \\
                    WB89-040 & 0.20(0.03) & 10.7 & -62.40(0.05) & 2.18(0.10) & 0.038 (0.004) &  & 0.30(0.09) & 10.7 & -62.00(0.05) & 1.8(0.2) &  & 1.3(0.2) & 10.7 & -62.40(0.05) & 1.90(0.12) &  & 1.3(0.2) & 10.7 & -62.50(0.06) & 1.40(0.14) \\
                    WB89-060 & ... & ... & ... & ... & ... &  & ... & ... & ... & ... &  & 1.4(0.3) & 10.7 & -83.70(0.14) & 3.2(0.3) &  & 4.8(0.6) & 10.7 & -83.90(0.03) & 2.30(0.07) \\
                    WB89-076 & 0.40(0.05) & 7.5 & -97.20(0.04) & 2.00(0.08) & 0.118 (0.007) &  & 1.0(0.2) & 7.5 & -97.60(0.03) & 1.70(0.09) &  & 1.8(0.3) & 4.9(0.4) & -97.10(0.03) & 1.40(0.07) &  & 2.1(0.3) & 7.5 & -97.10(0.03) & 1.40(0.07) \\
                    WB89-080 & 0.30(0.04) & 9.2 & -74.50(0.05) & 2.00(0.11) & 0.061 (0.005) &  & ... & ... & ... & ... &  & 1.0(0.3) & 9.2 & -74.1(0.2) & 3.0(0.6) &  & 0.8(0.2) & 9.2 & -74.2(0.2) & 1.5(0.4) \\
                    WB89-083 & 0.20(0.02) & 10.7 & -93.800(0.014) & 1.40(0.03) & 0.039 (0.001) &  & 1.4(0.2) & 10.7 & -93.90(0.02) & 1.60(0.04) &  & 0.68(0.10) & 10.7 & -93.80(0.03) & 1.20(0.07) &  & 0.40(0.09) & 10.7 & -93.60(0.04) & 0.80(0.09) \\
                    WB89-152 & 0.15(0.02) & 10.7 & -88.20(0.06) & 2.00(0.15) & 0.026 (0.003) &  & 1.3(0.2) & 10.7 & -88.60(0.04) & 3.00(0.09) &  & 0.5(0.2) & 10.7 & -88.50(0.12) & 0.8(0.3) &  & $\leq$ 0.6 & 10.7 & -88.5 & 1.8 \\
                    WB89-283 & 0.30(0.04) & 14.9 & -94.50(0.03) & 1.70(0.08) & 0.032 (0.002) &  & 1.0(0.2) & 14.9 & -94.40(0.01) & 1.40(0.04) &  & 0.7(0.2) & 7.7(1.1) & -94.70(0.07) & 2.0(0.2) &  & 0.30(0.07) & 14.9 & -94.70(0.07) & 1.1(0.2) \\
                    WB89-288 & 0.090(0.012) & 10.7 & -100.80(0.03) & 1.50(0.07) & 0.020 (0.001) &  & 1.20(0.15) & 10.7 & -100.900(0.008) & 1.70(0.02) &  & 0.53(0.10) & 10.7 & -101.00(0.07) & 1.7(0.2) &  & $\leq$ 0.3 & 10.7 & -101.0 & 1.8 \\
                    WB89-315 & 0.040(0.007) & 10.7 & -94.80(0.08) & 1.7(0.2) & 0.008 (0.001) &  & 0.50(0.07) & 10.7 & -94.90(0.04) & 2.07(0.10) &  & $\leq$ 0.2 & 10.7 &  & 2.1 &  & $\leq$ 0.4 & 10.7 & -95.1 & 1.8 \\
                    WB89-379 & 0.50(0.06) & 9.3 & -89.30(0.03) & 2.00(0.08) & 0.108 (0.006) &  & 0.7(0.2) & 9.3 & -89.50(0.03) & 1.40(0.12) &  & 0.70(0.11) & 6.4 & -89.20(0.06) & 1.90(0.14) &  & 1.0(0.2) & 9.3 & -89.10(0.05) & 1.60(0.12) \\
                    WB89-380 & 1.1(0.2) & 13.0 & -87.30(0.05) & 2.80(0.11) & 0.099 (0.003) &  & 5.1(0.7) & 13.0 & -87.40(0.05) & 2.60(0.11) &  & 1.6(1.3) & 11.0(8.0) & -87.4(0.3) & 2.8(0.8) &  & 2.3(0.5) & 13.0 & -86.1(0.2) & 4.8(0.5) \\
                    WB89-380-b & 0.30(0.07) & 11.4 & -84.70(0.09) & 1.8(0.2) & 0.047 (0.003) &  & 2.3(0.4) & 11.4 & -84.10(0.07) & 2.2(0.2) &  & 3.0(2.0) & 5.0(1.2) & -86.0(1.0) & 4.0(0.6) &  & 0.7(0.2) & 11.4 & -85.20(0.07) & 1.1(0.2) \\
                    WB89-391 & 0.50(0.06) & 9.7 & -86.10(0.03) & 1.70(0.07) & 0.114 (0.006) &  & 2.1(0.2) & 9.7 & -86.20(0.02) & 1.80(0.03) &  & 2.3(0.3) & 3.7 & -86.10(0.03) & 1.60(0.08) &  & 0.80(0.15) & 9.7 & -86.10(0.07) & 1.4(0.2) \\
                    WB89-399 & 0.50(0.06) & 11.0 & -82.00(0.02) & 1.80(0.05) & 0.092 (0.003) &  & 3.4(0.4) & 11.0 & -81.900(0.012) & 2.00(0.03) &  & 2.6(0.4) & 4.5(0.4) & -82.20(0.04) & 2.20(0.09) &  & $\leq$ 0.4 & 11.0 & 82.15 & 1.8 \\
                    WB89-437 & 2.6(0.3) & 11.8 & -71.50(0.04) & 2.80(0.08) & 0.270 (0.015) &  & 9.5(1.0) & 11.8 & -71.300(0.008) & 2.70(0.02) &  & 2.5(0.4) & 4.2 & -72.10(0.07) & 3.0(0.2) &  & 3.3(0.5) & 11.8 & -71.60(0.05) & 2.40(0.11) \\
                    WB89-440 & 0.20(0.02) & 10.7 & -72.00(0.02) & 1.70(0.04) & 0.042 (0.001) &  & 3.2(0.3) & 10.7 & -71.900(0.009) & 2.00(0.02) &  & 1.2(0.2) & 10.7 & -71.90(0.05) & 1.60(0.12) &  & $\leq$ 0.3 & 10.7 & -71.88 & 1.8 \\
                    WB89-501 & 0.70(0.09) & 11.0 & -58.40(0.03) & 2.10(0.07) & 0.107 (0.005) &  & 3.0(0.4) & 11.0 & -58.20(0.04) & 2.10(0.09) &  & 1.6(0.2) & 4.7 & -58.40(0.04) & 2.16(0.10) &  & 0.8(0.2) & 11.0 & -58.60(0.15) & 2.1(0.3) \\
                    WB89-529 & 0.11(0.02) & 10.7 & -59.80(0.04) & 1.60(0.09) & 0.023 (0.002) &  & 2.1(0.3) & 10.7 & -59.70(0.02) & 1.80(0.05) &  & 0.8(0.2) & 10.7 & -59.8(0.2) & 1.9(0.4) &  & $\leq$ 0.5 & 10.7 & -59.8 & 1.8 \\
                    WB89-572 & 0.20(0.03) & 10.7 & -47.70(0.02) & 2.00(0.04) & 0.041 (0.002) &  & 1.10(0.14) & 10.7 & -47.80(0.02) & 2.40(0.06) &  & 0.40(0.12) & 10.7 & -47.42(0.10) & 1.0(0.2) &  & 0.5(0.2) & 10.7 & -47.2(0.3) & 1.6(0.7) \\
                    WB89-621 & 1.4(0.2) & 10.3 & -25.40(0.04) & 2.60(0.08) & 0.0194 (0.010) &  & 4.0(0.4) & 10.3 & -25.63(0.01) & 3.10(0.03) &  & 0.80(0.13) & 6.7 & -25.70(0.06) & 2.00(0.15) &  & 2.6(0.4) & 10.3 & -25.10(0.04) & 2.22(0.10) \\
                    WB89-640 & 0.70(0.08) & 10.7 & -25.10(0.02) & 2.00(0.05) & 0.119 (0.006) &  & 1.1(0.3) & 10.7 & -25.10(0.03) & 1.10(0.09) &  & 2.3(0.4) & 10.7 & -24.90(0.09) & 2.4(0.2) &  & 3.9(0.7) & 10.7 & -25.40(0.09) & 2.6(0.2) \\
                    WB89-670 & 0.14(0.03) & 10.7 & -17.60(0.04) & 0.60(0.08) & 0.078 (0.011) &  & 1.2(0.2) & 10.7 & -17.68(0.01) & 0.90(0.02) &  & 1.8(0.4) & 7.8 & -17.70(0.04) & 0.67(0.10) &  & 2.1(0.6) & 10.7 & -17.70(0.06) & 0.60(0.12) \\
                    WB89-705 & 0.11(0.02) & 10.7 & -12.20(0.05) & 1.20(0.13) & 0.031 (0.005) &  & 0.60(0.11) & 10.7 & -12.20(0.05) & 1.40(0.12) &  & 1.3(0.2) & 8.2 & -12.200(0.013) & 0.70(0.03) &  & 1.9(0.3) & 10.7 & -12.20(0.02) & 0.60(0.04) \\
                    WB89-789 & 0.60(0.08) & 9.7 & 34.20(0.04) & 2.40(0.08) & 0.102 (0.005) &  & 3.9(0.4) & 9.7 & 34.200(0.015) & 3.10(0.04) &  & 1.4(0.3) & 7.1(1.0) & 34.30(0.06) & 1.90(0.15) &  & 3.3(0.5) & 9.7 & 34.20(0.05) & 2.00(0.12) \\
                    WB89-793 & 0.50(0.07) & 10.7 & 30.20(0.04) & 2.30(0.09) & 0.079 (0.006) &  & 2.3(0.3) & 10.7 & 29.80(0.02) & 1.70(0.05) &  & 1.4(0.4) & 10.7 & 30.5(0.2) & 2.1(0.4) &  & 2.5(0.4) & 10.7 & 30.20(0.06) & 1.8(0.2) \\
                    WB89-898 & 0.50(0.06) & 10.7 & 63.30(0.03) & 2.70(0.06) & 0.061 (0.003) &  & 2.2(0.3) & 10.7 & 63.30(0.03) & 3.20(0.07) &  & 1.2(0.3) & 10.7 & 63.50(0.14) & 2.6(0.3) &  & $\leq$ 0.3 & 10.7 & 63.5 & 1.8 \\
                    \noalign{\smallskip}
                    \hline
                \end{tabular}
                \tablefoot{
                    \tablefoottext{a}{Optical depth estimated from the hyperfine structure of HCN, by \textsc{MADCUBA}. The values are referring to the strongest line, i.e., HCN $J = 1-0, F = 2-1$ transition, at 88.63185~GHz, and they represent lower limit values due to the unknown source sizes and the assumed temperatures. The error shown is estimated by \textsc{MADCUBA}.}
                }
                \label{tab:abund1}
            \end{sidewaystable*}
            
            \begin{sidewaystable*}
                \centering
                \tiny
                \setlength{\tabcolsep}{1.75pt}
                \caption{Same as Table~\ref{tab:abund1} for HCO, SO, H$_2$CO, and CH$_3$OH.}
                    \begin{tabular}{lccccccccccccccccccccc}
                        \hline
                        \hline
                        \noalign{\smallskip}
                        &  \multicolumn{4}{c}{HCO} & \multicolumn{1}{c}{} & \multicolumn{4}{c}{SO} & \multicolumn{1}{c}{} & \multicolumn{5}{c}{H$_2$CO} & \multicolumn{1}{c}{} & \multicolumn{5}{c}{CH$_3$OH} \\
                        \noalign{\smallskip}
                        \cline{2-5} \cline{7-10} \cline{12-16} \cline{18-22}
                        \noalign{\smallskip}
                        Source & $N_\text{tot}$ & $T_\text{ex}$ & $V_\text{LSR}$ & FWHM & & $N_\text{tot}$ & $T_\text{ex}$ & $V_\text{LSR}$ & FWHM & & $N_\text{tot}$ & $T_\text{ex}$ & $V_\text{LSR}$ & FWHM & $\theta_\text{S}$ & & $N_\text{tot}$ & $T_\text{ex}$ & $V_\text{LSR}$ & FWHM & $\theta_\text{S}$ \\
                        & ($\times 10^{13}$~cm$^{-2}$) & (K) & (km s$^{-1}$) & (km s$^{-1}$) & & ($\times 10^{13}$~cm$^{-2}$) & (K) & (km s$^{-1}$) & (km s$^{-1}$) & & ($\times 10^{14}$~cm$^{-2}$) & (K) & (km s$^{-1}$) & (km s$^{-1}$) & ($''$) & & ($\times 10^{13}$~cm$^{-2}$) & (K) & (km s$^{-1}$) & (km s$^{-1}$) & ($''$) \\
                        \noalign{\smallskip}
                        \hline
                        \noalign{\smallskip}
                        19383+2711 & 0.40(0.12) & 10.7 & -69.7(0.3) & 3.9(0.6) &  & 0.32(0.10) & 10.7 & -69.9(0.5) & 3.6 & & 0.11(0.04) & 42.0(8.0) & -70.40(0.06) & 4.0(0.2) & ... &  & $\leq$ 0.1 & 10.7 & -70.2 & 1.9 & ...\\
                        19383+2711-b & 0.20(0.11) & 12.4 & -65.7(0.5) & 3.5(0.9) &  & 0.50(0.12) & 12.4 & -66.0(0.2) & 2.6(0.5) &  & 0.20(0.03) & 36.0(2.0) & -65.700(0.015) & 2.50(0.04) & ... &  & 1.3(0.2) & 12.4(0.8) & -65.90(0.04) & 2.0(0.1) & ...\\
                        19423+2541 & 0.40(0.06) & 9.9 & -72.50(0.08) & 3.1(0.2) &  & 2.4(0.3) & 9.9 & -73.00(0.06) & 3.10(0.14) &  & 0.60(0.12) & 41.0(3.0) & -72.60(0.02) & 4.00(0.05) & 35 &  & 4.6(0.7) & 9.9(0.4) & -72.40(0.06) & 4.00(0.13) & 28 \\
                        19489+3030 & $\leq$ 0.1 & 10.7 & -69.29 & 2.1 &  & 0.30(0.11) & 10.7 & -69.5(0.2) & 1.8(0.6) &  & ... & ... & ... & ... & ... &  & ... & ... & ... & ... & ...\\
                        19571+3113 & 0.30(0.12) & 10.7 & -65.5(0.6) & 4.8(1.3) &  & $\leq$ 0.1 & 10.7 & -66.2 & 2.0 &  & ... & ... & ... & ... & ... &  & ... & ... & ... & ... & ... \\
                        19571+3113-b & 0.30(0.11) & 10.7 & -61.1(0.4) & 3.4(0.7) &  & $\leq$ 0.1 & 10.7 & -61.7 & 2.0 &  & ... & ... & ... & ... & ... &  & ... & ... & ... & ... & ... \\
                        20243+3853 & 0.30(0.05) & 10.7 & -73.10(0.06) & 2.30(0.13) &  & 0.50(0.13) & 10.7 & -73.5(0.2) & 3.0(0.5) &  & ... & ... & ... & ... & ... &  & ... & ... & ... & ... & ...  \\
                        WB89-002 & 0.20(0.05) & 10.7 & -2.80(0.06) & 0.90(0.15) &  & $\leq$ 0.3 & 10.7 & -2.83 & 2.0 &  & ... & ... & ... & ... & ... &  & ... & ... & ... & ... & ...  \\
                        WB89-006 & $\leq$ 0.1 & 9.8 & -92.3 & 2.1 &  & 0.24(0.10) & 9.8 & -92.0 & 2.0 &  & 0.053(0.010) & 29.0(2.0) & -92.10(0.05) & 1.70(0.07) & ... &  & 1.1(0.3) & 9.8(0.7) & -92.00(0.09) & 1.6(0.2) & ... \\
                        WB89-006-b & 0.07(0.04) & 9.4 & -90.5(0.3) & 1.2(0.7) &  & $\leq$ 0.2 & 9.4 & -90.3 & 2.0 &  & 0.040(0.008) & 23.0(2.0) & -90.30(0.06) & 1.92(0.10) & ... &  & 0.7(0.2) & 9.4(1.0) & -90.30(0.09) & 1.3(0.2) & ... \\
                        WB89-014 & $\leq$ 0.1 & 10.7 & -96.0 & 2.1 &  & $\leq$ 0.2 & 10.7 & -96.0 & 2.0 &  & ... & ... & ... & ... & ... &  & ... & ... & ... & ... & ... \\
                        WB89-031 & 0.12(0.03) & 10.7 & -88.8(0.2) & 1.8(0.4) &  & $\leq$ 0.1 & 10.7 & -88.89 & 2.0 &  & ... & ... & ... & ... & ... &  & ... & ... & ... & ... & ... \\
                        WB89-035 & 0.14(0.03) & 11.7 & -77.60(0.08) & 1.3(0.2) &  & 0.80(0.13) & 11.7 & -78.10(0.05) & 1.70(0.12) &  & 0.14(0.03) & 32.0(3.0) & -77.60(0.02) & 2.40(0.05) & 32 &  & 1.2(0.2) & 12.0(1.3) & -77.70(0.06) & 2.40(0.14) & 35 \\
                        WB89-040 & 0.30(0.05) & 10.7 & -62.30(0.08) & 2.1(0.2) &  & 0.70(0.14) & 10.7 & -62.20(0.08) & 1.6(0.2) &  & ... & ... & ... & ... & ... &  & ... & ... & ... & ... & ... \\
                        WB89-060 & $\leq$ 0.1 & 10.7 & -83.7 & 2.2 &  & 3.6(0.5) & 10.7 & -84.00(0.05) & 2.50(0.12) &  & ... & ... & ... & ... & ... &  & ... & ... & ... & ... & ... \\
                        WB89-076 & 0.06(0.02) & 7.5 & -97.00(0.15) & 1.3(0.3) &  & 1.7(0.2) & 7.5 & -97.30(0.02) & 1.30(0.05) & & 0.060(0.011) & 28.0(3.0) & -97.30(0.02) & 1.90(0.05) & 98 & & 1.7(0.5) & 7.5(1.1) & -97.30(0.06) & 1.50(0.14) & 33 \\
                        WB89-080 & 0.07(0.02) & 9.2 & -73.90(0.09) & 0.7 &  & 0.80(0.14) & 9.2 & -74.40(0.04) & 1.00(0.09) &  & 0.12(0.03) & 30.0(5.0) & -74.50(0.03) & 1.70(0.07) & 32 &  & 1.4(0.3) & 9.2(0.9) & -74.40(0.06) & 1.90(0.14) & 32 \\
                        WB89-083 & 0.10(0.03) & 10.7 & -93.70(0.11) & 1.4(0.3) &  & $\leq$ 0.1 & 10.7 & -93.76 & 2.0 &  & ... & ... & ... & ... & ... &  & ... & ... & ... & ... & ... \\
                        WB89-152 & $\leq$ 0.1 & 10.7 & -88.5 & 2.1 &  & $\leq$ 0.3 & 10.7 & -88.5 & 2.0 &  & ... & ... & ... & ... & ... &  & ... & ... & ... & ... & ... \\
                        WB89-283 & 0.20(0.05) & 14.9 & -94.50(0.14) & 2.6(0.3) &  & 0.30(0.09) & 14.9 & -94.7(0.3) & 2.5(0.7) &  & 0.12(0.02) & 35.0(4.0) & -94.50(0.02) & 1.60(0.04) & 36 &  & 1.1(0.2) & 15.0(1.1) & -94.40(0.03) & 1.40(0.07) & 32 \\
                        WB89-288 & 0.20(0.04) & 10.7 & -100.60(0.07) & 1.4(0.2) &  & $\leq$ 0.2 & 10.7 & -101.1 & 2.1 &  & ... & ... & ... & ... & ... &  & ... & ... & ... & ... & ... \\
                        WB89-315 & $\leq$ 0.1 & 10.7 & -95.1 & 2.1 &  & $\leq$ 0.2 & 10.7 & -95.1 & 2.0 &  & ... & ... & ... & ... & ... &  & ... & ... & ... & ... & ... \\
                        WB89-379 & 0.20(0.03) & 9.3 & -89.20(0.09) & 2.3(0.2) &  & 0.4(0.1) & 9.3 & -89.20(0.11) & 1.5(0.3) &  & 0.13(0.02) & 33.0(2.0) & -89.400(0.011) & 1.80(0.03) & 34 &  & 1.6(0.3) & 9.3(0.9) & -89.30(0.05) & 1.70(0.11) & 30\\
                        WB89-380 & 0.60(0.08) & 13.0 & -86.60(0.06) & 2.90(0.13) &  & 1.1(0.2) & 13.0 & -86.7(0.2) & 4.1(0.4) &  & 0.20(0.07) & 39.0(8.0) & -87.6(0.2) & 2.3(0.2) & ... &  & 0.20(0.13) & 13.0(4.0) & -88.4(0.1) & 1.0(0.3) & ... \\
                        WB89-380-b & $\leq$ 0.1 & 11.4 & -86.0 & 2.1 &  & 0.40(0.12) & 11.4 & -85.20(0.06) & 1.0(0.2) &  & 0.15(0.06) & 22.0(2.0) & -85.6(0.3) & 2.8(0.3) & ... &  & 2.5(0.4) & 11.4(0.5) & -86.30(0.08) & 3.0(0.2) & ... \\
                        WB89-391 & 0.30(0.04) & 9.7 & -85.90(0.04) & 1.71(0.10) &  & 1.10(0.15) & 9.7 & -86.00(0.03) & 1.30(0.06) &  & 0.090(0.015) & 25.0(2.0) & -86.00(0.02) & 1.60(0.04) & 49 &  & 1.7(0.3) & 9.7(0.8) & -85.90(0.03) & 1.40(0.07) & 29\\
                        WB89-399 & 0.40(0.07) & 11.0 & -81.80(0.09) & 2.0(0.2) &  & $\leq$ 0.2 & 11.0 & -82.15 & 2.0 &  & 0.11(0.03) & 46.0(7.0) & -82.20(0.02) & 1.60(0.04) & 128 &  & 0.30(0.09) & 11.0(3.0) & -82.20(0.05) & 1.00(0.13) & 56 \\
                        WB89-437 & 0.20(0.05) & 11.8 & -71.8(0.2) & 2.9(0.4) &  & 3.6(0.5) & 11.8 & -71.60(0.04) & 2.81(0.10) &  & 1.00(0.15) & 33.0(2.0) & -71.40(0.02) & 2.80(0.04) & 21 &  & 11.0(1.4) & 11.8(0.4) & -71.60(0.03) & 3.00(0.07) & 22 \\
                        WB89-440 & 0.40(0.06) & 10.7 & -71.70(0.04) & 1.66(0.10) &  & $\leq$ 0.2 & 10.7 & -71.88 & 2.0 &  & ... & ... & ... & ... & ... &  & ... & ... & ... & ... & ...  \\
                        WB89-501 & 0.30(0.05) & 11.0 & -58.30(0.07) & 2.1(0.2) &  & 0.60(0.11) & 11.0 & -58.50(0.06) & 1.90(0.15) &  & 0.20(0.04) & 34.0(3.0) & -58.50(0.02) & 2.00(0.04) & 30 &  & 1.9(0.4) & 11.0(1.2) & -58.60(0.05) & 1.70(0.11) & 31 \\
                        WB89-529 & 0.30(0.05) & 10.7 & -59.50(0.07) & 1.3(0.2) &  & $\leq$ 0.3 & 10.7 & -59.8 & 2.0 &  & ... & ... & ... & ... & ... &  & ... & ... & ... & ... & ... \\
                        WB89-572 & $\leq$ 0.1 & 10.7 & -47.4 & 2.2 &  & 0.7(0.2) & 10.7 & -47.9(0.1) & 1.7(0.2) &  & ... & ... & ... & ... & ... &  & ... & ... & ... & ... & ...  \\
                        WB89-621 & 0.14(0.03) & 10.3 & -25.70(0.14) & 2.2(0.3) &  & 5.2(0.7) & 10.3 & -25.30(0.03) & 2.00(0.07) &  & 0.40(0.07) & 25.0(2.0) & -25.30(0.03) & 2.40(0.06) & 38 &  & 10.0(1.4) & 10.3(0.4) & -25.40(0.02) & 1.70(0.05) & 26\\
                        WB89-640 & 0.80(0.11) & 10.7 & -25.10(0.06) & 2.50(0.14) &  & 1.0(0.2) & 10.7 & -25.20(0.12) & 1.8(0.3) &  & ... & ... & ... & ... & ... &  & ... & ... & ... & ... & ...\\
                        WB89-670 & $\leq$ 0.1 & 10.7 & -17.65 & 2.1 &  & $\leq$ 0.2 & 10.7 & -17.65 & 2.0 &  & ... & ... & ... & ... & ... &  & ... & ... & ... & ... & ...  \\
                        WB89-705 & 0.07(0.02) & 10.7 & -11.90(0.08) & 0.7(0.2) &  & 0.50(0.07) & 10.7 & -12.20(0.02) & 0.60(0.04) &  & ... & ... & ... & ... & ... &  & ... & ... & ... & ... & ... \\
                        WB89-789 & 0.30(0.06) & 9.7 & 34.20(0.07) & 1.9(0.2) &  & 0.9(0.2) & 9.7 & 34.1(0.2) & 2.8(0.4) &  & 0.30(0.11) & 44.0(8.0) & 34.20(0.03) & 2.10(0.07) & 22 &  & 1.5(0.3) & 9.7(1.0) & 34.10(0.05) & 2.00(0.13) & 32\\
                        WB89-793 & $\leq$ 0.1 & 10.7 & 30.5 & 2.1 &  & 1.2(0.3) & 10.7 & 30.20(0.07) & 1.3(0.2) &  & ... & ... & ... & ... & ... &  & ... & ... & ... & ... & ...  \\
                        WB89-898 & 0.12(0.04) & 10.7 & 63.4(0.3) & 2.0 &  & 0.8(0.2) & 10.7 & 63.2(0.2) & 2.5(0.5) &  & ... & ... & ... & ... & ... &  & ... & ... & ... & ... & ... \\
                        \noalign{\smallskip}
                        \hline
                    \end{tabular}
                \label{tab:abund2}
            \end{sidewaystable*}


        \section{Molecular column densities}
            \label{app:densities}
\begin{figure*}
                \centering
                \begin{minipage}{0.49\textwidth}
                    \centering
                    \includegraphics[width=\textwidth]{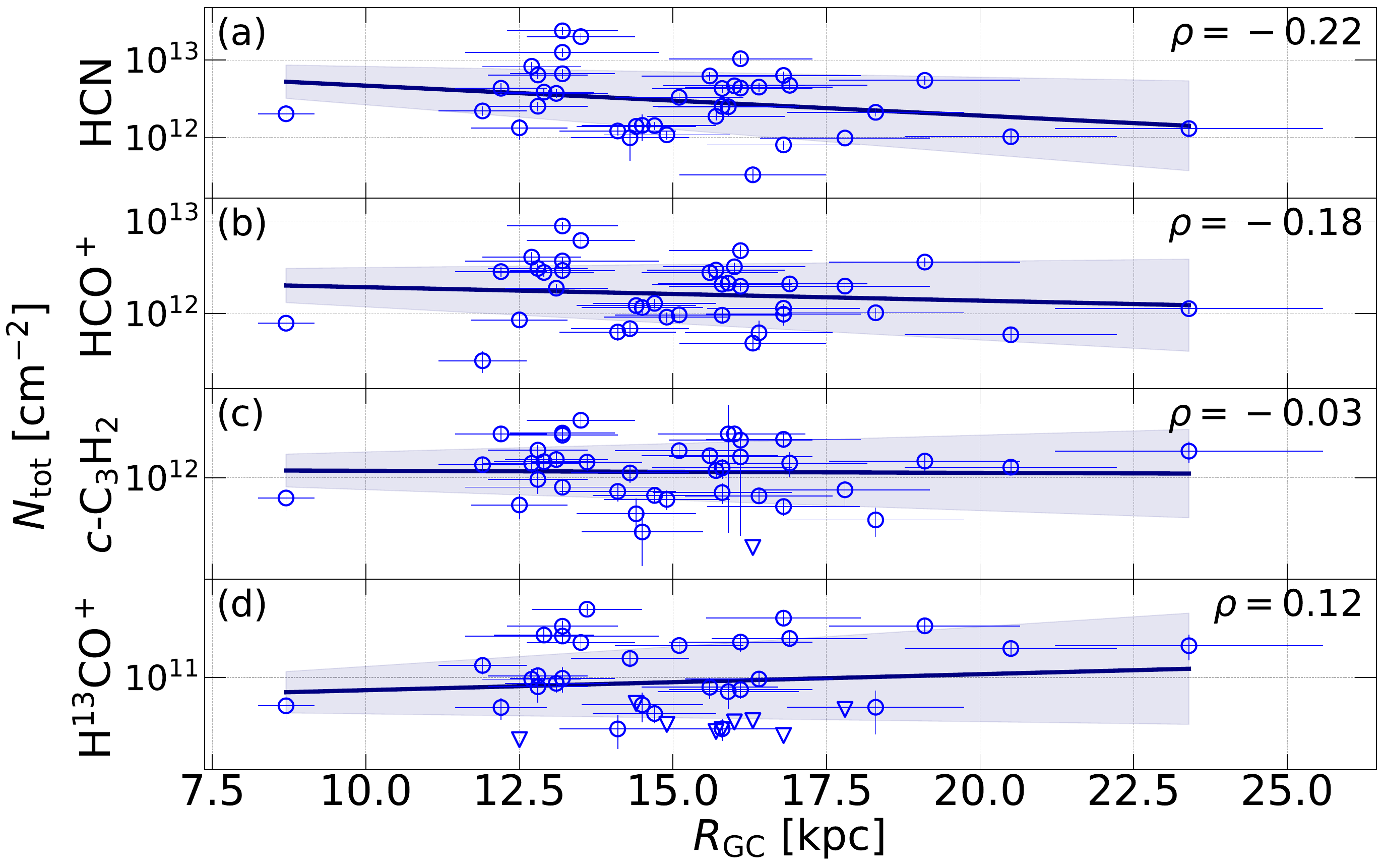}
                \end{minipage}
                \hfill
                \begin{minipage}{0.495\textwidth}
                    \centering
                    \includegraphics[width=\textwidth]{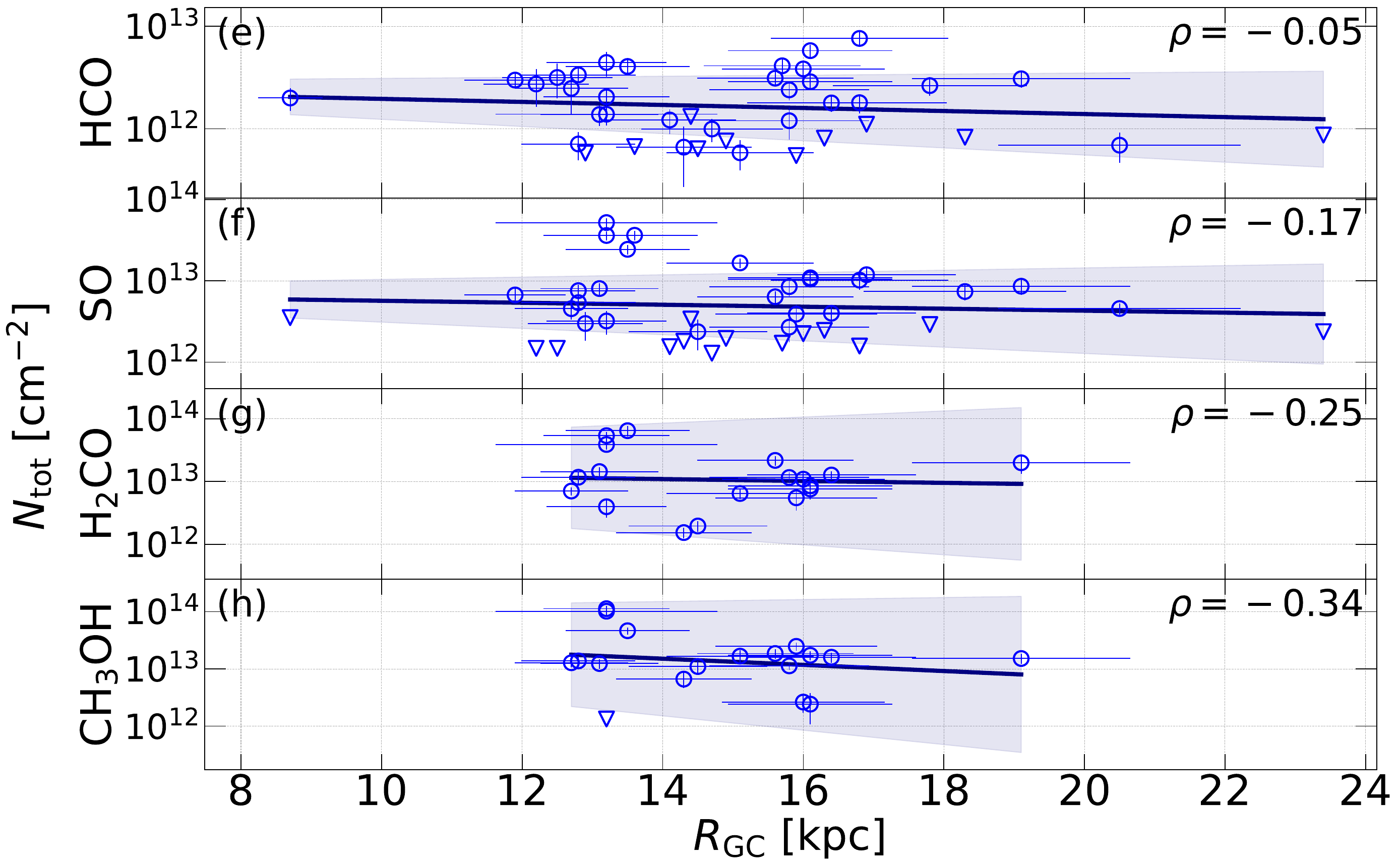}
                \end{minipage}
                \caption{Galactocentric gradients of column densities, $N_\text{tot}$. The plots show the trends for HCN (a), HCO$^+$ (b), \textit{c}-C$_3$H$_2$ (c), H$^{13}$CO$^+$ (d), HCO (e), SO (f), H$_2$CO (g), and CH$_3$OH (h) as a function of galactocentric radius ($R_\text{GC}$). The upper limit values are represented with triangles. The dark blue lines represent the linear regressions computed over the datasets. In the upper-right side of each subplot, the Pearson correlation coefficient, $\rho$, is shown.}
                \label{fig:densities}
            \end{figure*}

            The derived column densities of the 39 molecular clumps for the studied molecules are within the ranges shown in Table~\ref{tab:ranges}.        
            The trends of $N_\text{tot}$ as a function of the galactocentric radius for all molecules are shown in Fig.~\ref{fig:densities}. 
            A linear fit is performed for each plot, and the Pearson correlation coefficient, $\rho$, is calculated.
            As illustrated, almost constant trends and no- or weak correlations (i.e. $\left| \rho \right| \lesssim 0.2$) are found for HCN, HCO$^+$, \textit{c}-C$_3$H$_2$, H$^{13}$CO$^+$, HCO, SO, and H$_2$CO. A higher anti-correlation is found between CH$_3$OH and the galactocentric distance.

        \section{Galactocentric gradients of H$^{13}$CN}
            \label{app:isotop}
\begin{figure*}
                \centering
                \includegraphics[width=\linewidth]{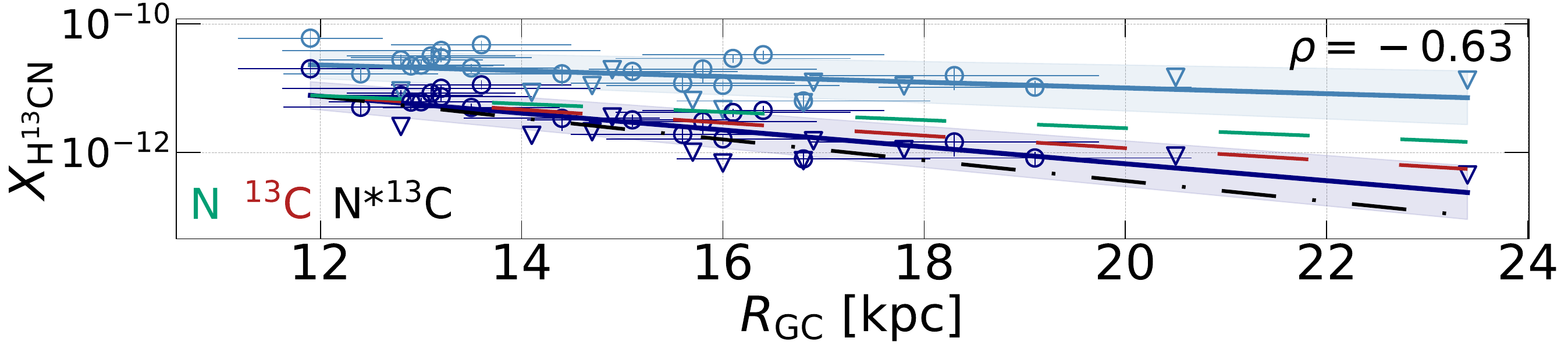}
                \caption{Galactocentric gradients of fractional abundances w.r.t. H$_2$, $X_\text{mol}$, of H$^{13}$CN. The light blue data represent the abundances calculated using a constant gas-to-dust ratio ($\gamma=100$), while the dark blue data illustrate those estimated using a non-constant gas-to-dust ratio \citep[$\gamma = \gamma(R_\text{GC})$ from][]{giannetti2017galactocentric}. For both datasets, the linear regression results are shown as the light blue and dark blue lines, respectively. The $1\sigma$ error bars over the slope of the gradients are displayed for each molecular fit. The upper limit values are represented with triangles. The gradients of the elemental abundances of carbon ($^{13}$C) and nitrogen (N), as reported by \citet{mendez2022gradients}, are plotted as dashed lines. The product of the parent elements of the species is represented by a black dash–dot line. All the elemental trends are plotted shifted to the start of the molecular gradients estimated with $\gamma(R_\text{GC})$ for reference to allow comparison of their slopes with the linear fit of the molecular abundances. In the upper-right side of each plot, the Pearson correlation coefficient, $\rho$, is shown (estimated only for the abundances estimated using the non-constant gas-to-dust ratio).}
                \label{fig:abundances2}
            \end{figure*}

            The galactocentric gradients of H$^{13}$CN fractional abundances are presented in Fig.~\ref{fig:abundances2}. This isotologue of HCN is characterized by a decreasing rate of $\sim114$, between the Solar circle and 24~kpc, higher than that of $^{13}$C (and N). A steeper decrease of this species suggests a lower formation efficiency in the OG, with respect to the inner Galaxy. It is important to highlight the contrasting behavior between HCN and HCO$^+$, compared to their $^{13}$C isotopologues: while HCO$^+$ shows a steeper decrease than H$^{13}$CO$^+$, HCN displays a shallower gradient compared to its $^{13}$C-bearing isotopologue.

\section{Correlation between fractional abundances and between line widths}
\begin{figure*}
        \centering
        \begin{subfigure}{0.48\textwidth}
            \includegraphics[width=\linewidth]{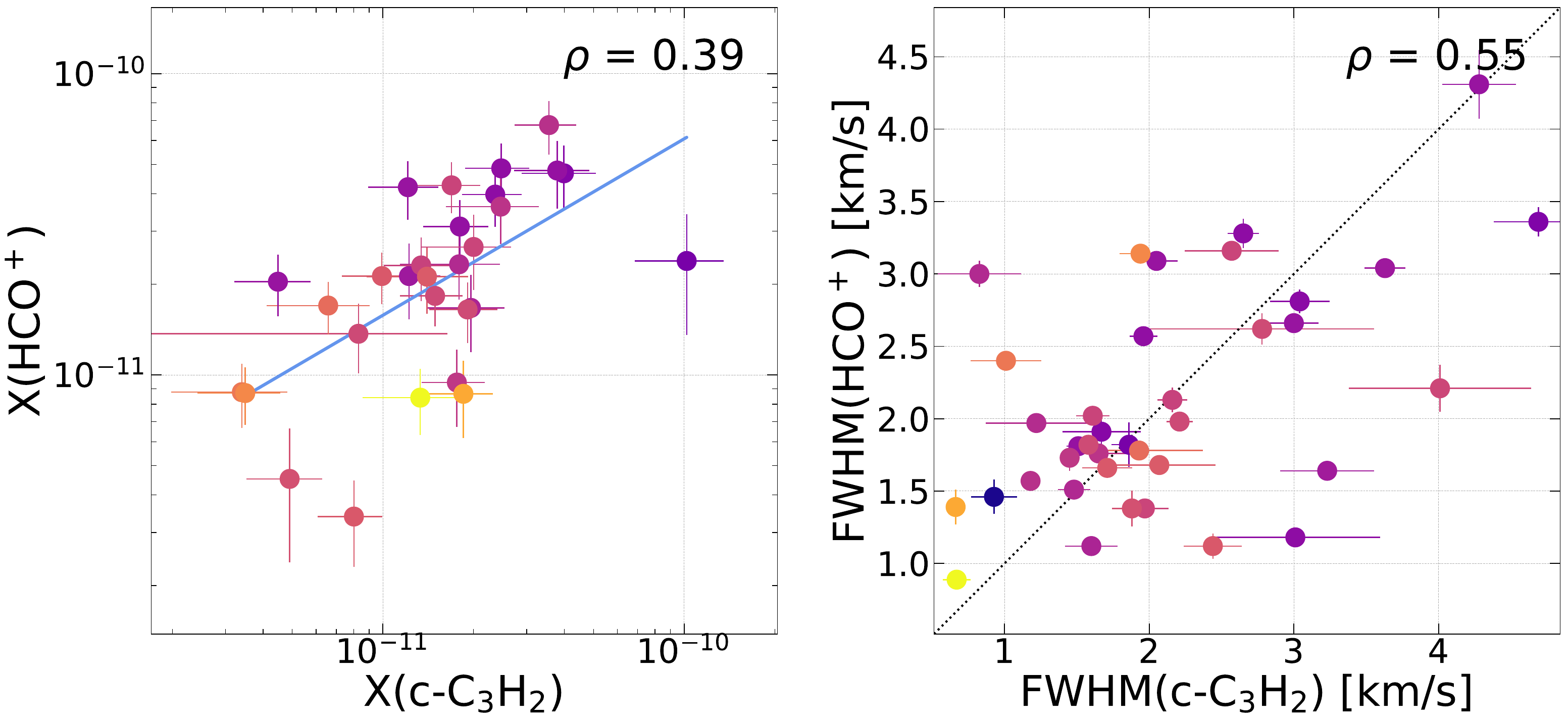}           
        \end{subfigure}
        \hfill
        \begin{subfigure}{0.48\textwidth}
            \includegraphics[width=\linewidth]{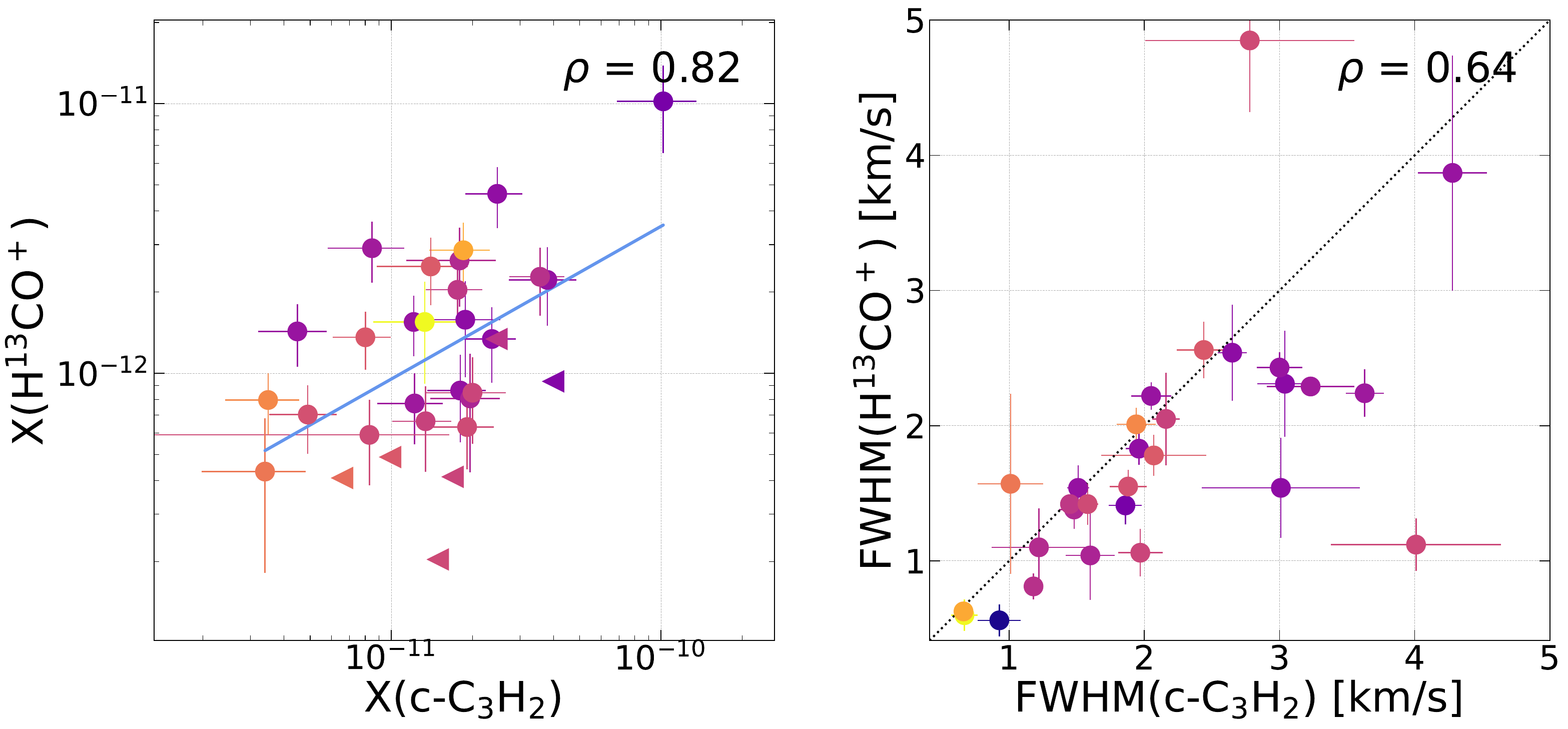}
        \end{subfigure}
    
        \vspace{1em}

        \begin{subfigure}{0.48\textwidth}
            \includegraphics[width=\linewidth]{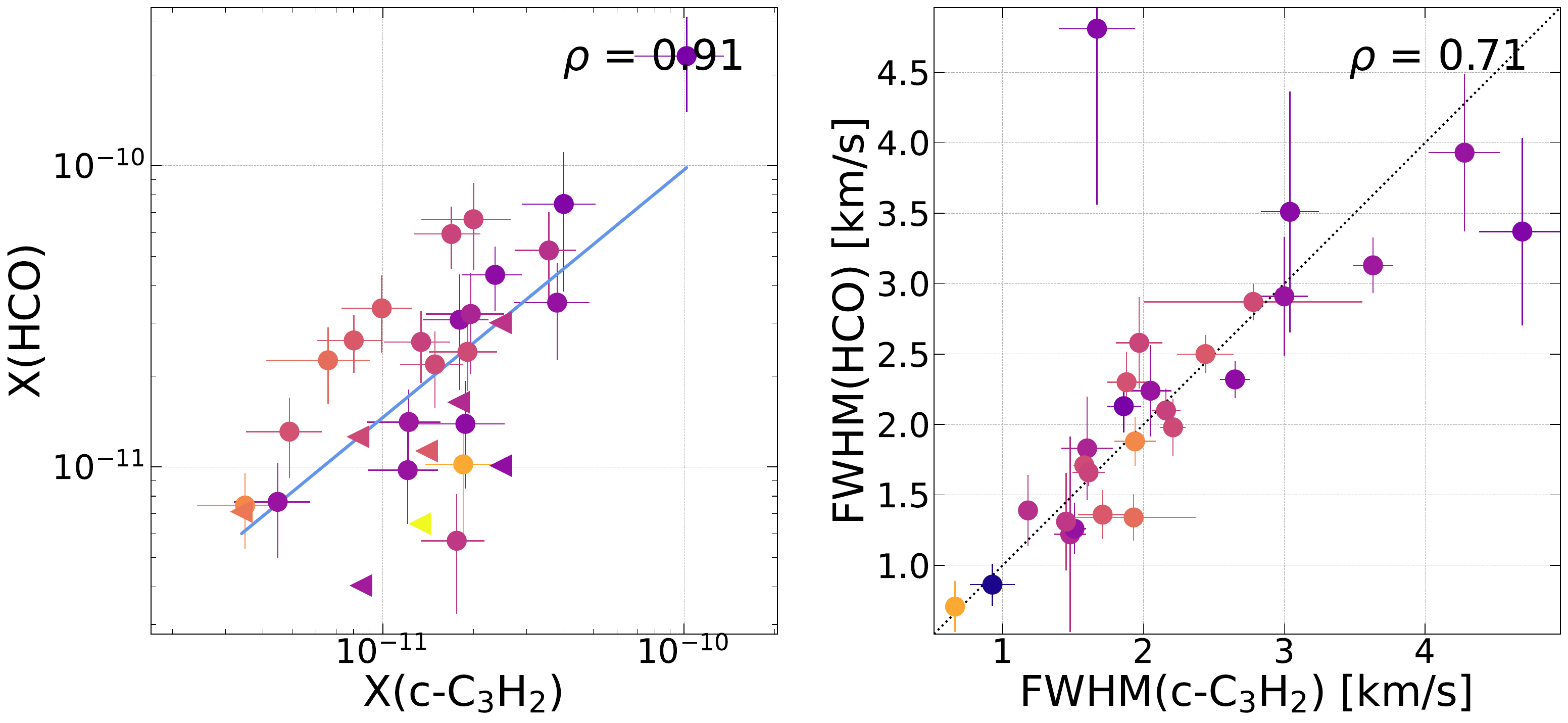}
        \end{subfigure}
        \hfill
        \begin{subfigure}{0.48\textwidth}
            \includegraphics[width=\linewidth]{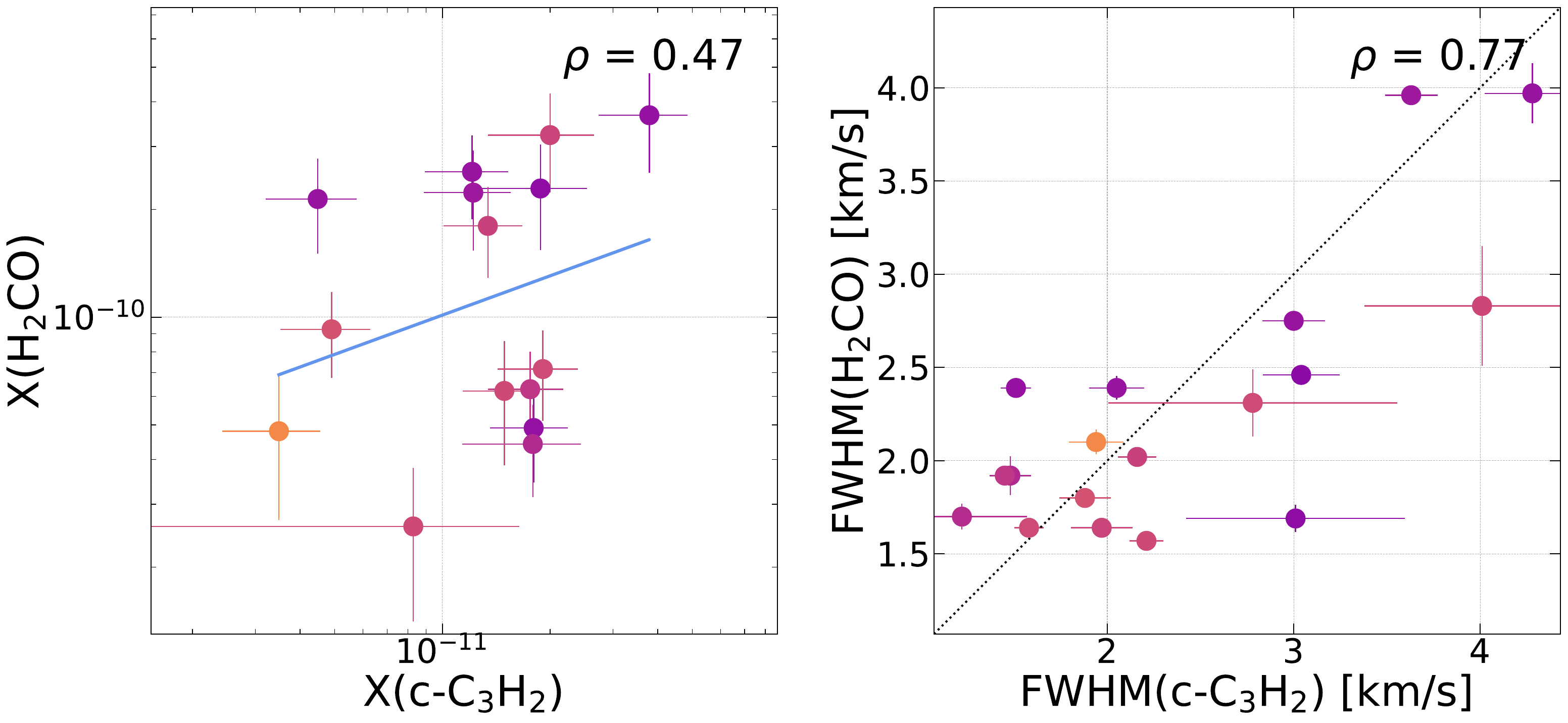}
        \end{subfigure}
    
        \vspace{1em}
    
        \begin{subfigure}{0.48\textwidth}
            \includegraphics[width=\linewidth]{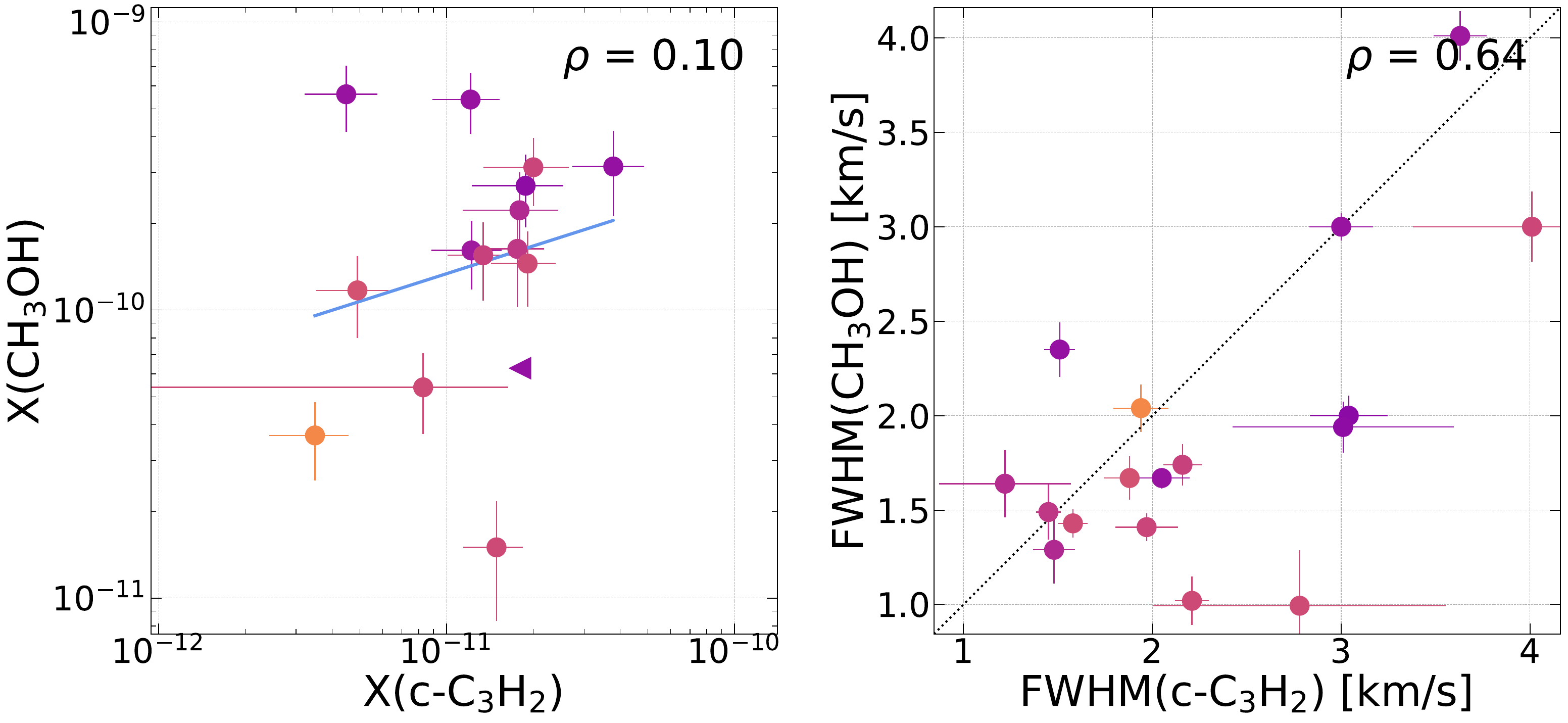}
        \end{subfigure}
    
        \vspace{1em}

        \begin{subfigure}{\textwidth}
            \centering
            \includegraphics[width=0.9\linewidth]{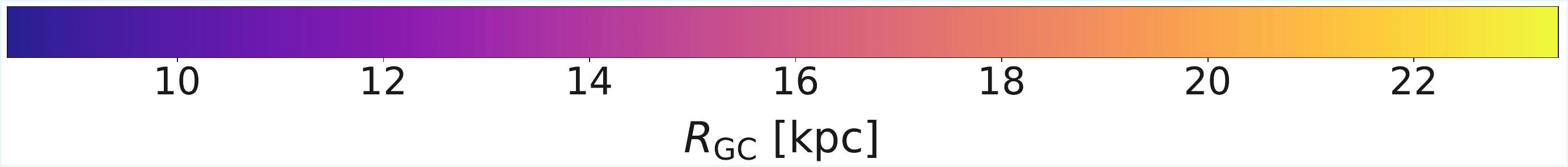}
        \end{subfigure}
    
        \caption{Correlation plots between fractional abundances and between line widths, of CO-related species (i.e., HCO$^+$, H$^{13}$CO$^+$, HCO, H$_2$CO, and CH$_3$OH) and \textit{c}-C$_3$H$_2$. A linear fit is performed for the abundance plots. The upper limit values are represented with triangles. The Pearson correlation coefficient, $\rho$, is estimated and shown in all the plots (top-right). Galactocentric radius is color-coded for each source with the color scale on the bottom.}
        \label{fig:correlations}
    \end{figure*}

    Figure~\ref{fig:correlations} shows plots to highlight possible correlations between fractional abundances and between line widths of different species. In particular, we analyzed the correlation between HCO$^+$, H$^{13}$CO$^+$, HCO, H$_2$CO, CH$_3$OH (i.e., CO-product molecules), and \textit{c}-C$_3$H$_2$, produced from the "free" carbon. 
    
    We found a high correlation between the abundances of \textit{c}-C$_3$H$_2$ with HCO and H$^{13}$CO$^+$, with $\rho \sim 0.8 - 0.9$, and a good correlation between their line width, indicating how these two couples of molecules trace the same gas. The correlations with CH$_3$OH and H$_2$CO are less constrained, for both abundances and line widths. As said in Sect.~\ref{dis:ratios}, we would expect a low correlation coefficient between the species chemically linked to the CO and \textit{c}-C$_3$H$_2$ because of their different formation pathways, and their weak chemical link. The correlation between these species may not result from a direct chemical link. Still, it could arise from their simultaneous formation in the star-forming region or from the similar gas they trace. 

\end{appendix}
    
\end{document}